\documentclass[journal,twoside]{IEEEtran} 

\usepackage{multirow}
\usepackage{booktabs}
\usepackage{tabularx}
\usepackage{ragged2e}
\usepackage[final]{microtype}
\newcolumntype{L}{>{\RaggedRight\arraybackslash}X} 

\usepackage[symbol]{footmisc}

\usepackage{cite}

\ifCLASSINFOpdf
  \usepackage[pdftex]{graphicx}
\else
  \usepackage[dvips]{graphicx}
\fi

\usepackage{amsmath}
\usepackage{amssymb,amsfonts,comment}
\usepackage[alpha]{mdpn} 

\usepackage[noend]{algpseudocode}
\usepackage[noend,ruled,vlined,linesnumbered]{algorithm2e}

\usepackage{array}

\ifCLASSOPTIONcompsoc
  \usepackage[caption=false,font=normalsize,labelfont=sf,textfont=sf]{subfig}
\else
  \usepackage[caption=false,font=footnotesize]{subfig}
\fi

\begin{document}

\title{A Heuristically Assisted Deep Reinforcement Learning Approach for Network Slice Placement}

\author{José~Jurandir~Alves~Esteves,~\IEEEmembership{Student Member,~IEEE,} Amina~Boubendir,~\IEEEmembership{Member,~IEEE,} Fabrice~Guillemin, and~Pierre~Sens,~\IEEEmembership{Member,~IEEE}


\thanks{J.J. Alves Esteves is with Orange Labs, 92320 Chatilon, France and also with LIP6 -- Inria, Sorbonne Univ., 75005 Paris, France (e-mail: josejurandir.alvesesteves@orange.com).}
\thanks{A. Boubendir and F. Guillemin are with Orange Labs, 92320 Chatillon, France (e-mail: firstname.name@orange.com).}
\thanks{P. Sens is with LIP6 -- Inria, Sorbonne Univ., CNRS, 75005 Paris, France (e-mail: pierre.sens@lip6.fr).}
}

\markboth{IEEE TRANSACTIONS ON NETWORK AND SERVICE MANAGEMENT}%
{ALVES ESTEVES \MakeLowercase{\textit{et al.}}: A HEURISTICALLY ASSISTED DEEP REINFORCEMENT LEARNING APPROACH FOR NETWORK SLICE PLACEMENT}


\maketitle

\begin{abstract}
Network Slice placement with the problem of allocation of resources from a virtualized substrate network is an optimization problem which can be formulated as a multi-objective Integer Linear Programming (ILP) problem. However, to cope with the complexity of such a continuous task and seeking for optimality and automation, the use of  Machine Learning (ML) techniques appear as a promising approach. We introduce a hybrid placement solution based on Deep Reinforcement Learning (DRL) and a dedicated optimization heuristic based on the "Power of Two Choices" principle. The DRL algorithm uses the so-called Asynchronous Advantage Actor Critic (A3C) algorithm for fast learning, and Graph Convolutional Networks (GCN) to automate feature extraction from the physical substrate network. The proposed Heuristically-Assisted DRL (HA-DRL) allows to accelerate the learning process and gain in resource usage when compared against other state-of-the-art approaches as the evaluation results evidence.
\end{abstract}


\ifCLASSOPTIONpeerreview
\begin{center} \bfseries EDICS Category: 3-BBND \end{center}
\fi


\IEEEpeerreviewmaketitle

\section{Introduction}

\IEEEPARstart{N}{etwork} Slicing is a major stake in 5G networks, which is notably enabled by virtualization techniques applied to Network Functions---a.k.a. Network Function Virtualization (NFV)---and by Software Defined Network (SDN) techniques \cite{NFV}. Thanks to these technical enablers now widely used in the telecom industry, a telecommunications network becomes a programmable platform, which can offer virtual networks enriched by Virtual Network Functions (VNFs) and IT resources, and can be tailored to the specific needs of certain customers (e.g., companies) or vertical markets (automotive, e-health, etc.). These augmented virtual networks give rise to the concept of Network Slicing which is specified by standardization bodies \cite{3GPP,etsi}. In this paper, we shall consider a Network Slice as a set of VNFs interconnected by a transport network and with networking (bandwidth) and computing (CPU, RAM) requirements. The Network Slice Placement problem can be viewed as a variant of Virtual Network Embedding (VNE) or VNF Forwarding Graph Embedding (VNF-FGE) and there is a huge amount of literature on this topic \cite{vne_alevin, vne_survey,survey_vnf_ra_2016,survey_vfnp}. Such an optimization problem is usually modeled as an Integer Linear Programming (ILP) problem which turns out to be $\mathcal{NP}$-hard \cite{vne_np_hardness} with very long convergence time.

Therefore, heuristics have been developed (see \cite{cnsm_2020} for an extensive list of heuristics for slice placement). For instance, in \cite{cnsm_2020}, a heuristic based on the "Power of two Choices" principle (for short, P2C) is proposed and gives satisfactory results, both in terms of convergence time and slice acceptance ratio. From an operational perspective, heuristic approaches are more suitable than ILP as they yield faster placement results. This is very important for operational networks because traffic conditions are fluctuating and placement response time is an important performance indicator in the customer relationship. The drawback of heuristic approaches is that they give sub-optimal solutions. To remedy  this problem, Machine Learning (ML) offer a corpus of methods able to overcome the convergence issues of ILPs while being more accurate than heuristics. Deep Reinforcement Learning (DRL) has recently been used in the context of VNE and VNF-FGE \cite{p1,p2,p5}.

A DRL agent is theoretically capable of learning an optimal decision policy only based on its own experience; this property eliminates the need for an accurate training data set that may not be available. However, from a practical point of view, ensuring that the DRL agent converges to an optimal policy is a challenge since the agent acts as a self-controlled black box. In addition, there are a large number of hyper-parameters to fine-tune to ensure an adequate equilibrium between exploring solutions and exploiting the knowledge acquired via training. While there are techniques to improve the efficiency of the solution exploration process (e.g., $\epsilon$-greedy, entropy regularization), their use may also lead to situations of instability, where the algorithm may diverge from the optimal point. 

To overcome this unsuitable behavior of DRL agents, we introduce in the present paper the concept of Heuristically Assisted DRL (HA-DRL) to accelerate and stabilize the convergence of DRL techniques when applied to the Network Slice Placement. The proposed contributions are twofold: i) We propose a DRL algorithm combining Advantage Actor Critic and a Graph Convolutional Network (GCN) to solve Network Slice Placement optimization problem \cite{p1}; ii) We reinforce the DRL learning process by using the P2C based heuristic \cite{cnsm_2020} to control the DRL convergence.

The organization of this paper is as follows: In Section~\ref{sec:sota}, we review related work on slice placement and more generally on VNE by paying special attention to ML techniques. In Section~\ref{sec:network_model}, we describe the Network Slice Placement problem and introduce the various elements of the model. The multi-objective optimization problem of Network Slice Placement is formulated in Section~\ref{sec:problem_statement_and_formulation}. \\

A DRL approach to solving the multi-objective optimization problem  is described in Section~\ref{sec:drl_proposal}. The control of the DRL convergence by using a the P2C heuristic is introduced in Section~\ref{sec:aidedDRL}. The experiments and evaluation results are presented in Section~\ref{sec:evaluation}, while conclusions and perspectives are presented  in Section~\ref{sec:conclusion}.

\section{Related Work Analysis \label{sec:sota}}

We review, in this section, recent studies on the network slice placement problem. We consider comprehensive surveys such as \cite{vne_alevin, vne_survey,survey_vnf_ra_2016,survey_vfnp,p1} and analyse existing works along two lines: i) ML-based approaches for slice placement optimization (Section~\ref{sec:ml_based}), and ii) hybrid approaches combining both heuristics and ML for slice placement (Section~\ref{sec:hybrid}). 

\subsection{On ML Approaches for Slice Placement Optimization \label{sec:ml_based}}

The most relevant ML-based approaches for network slice placement optimization are DRL-based solutions. We analyze: i) the RL setup elements of each solution (i.e., state, action, reward); and ii) the modeling aspects of DRL (i.e., type of convolutions used, training algorithms used). 

\subsubsection{RL Setup Elements}
\paragraph{State representation}
The state representation is almost similar in all analyzed solutions. In \cite{p1, rkhami2021learn, p2, p5}, only resource-related features are used to represent the state: i) the number of resources available on physical nodes (CPU and RAM) and links (bandwidth) and ii) the number of these resources required by the VNFs and Virtual Links (VLs) of the network slice to be placed. Latency-related features are also considered in \cite{p3, p4, quang2019deep, p8}. However, adding such features to the state brings additional complexity to their models. In particular, the model in \cite{p3} requires additional chaining information and performance indexes for the VNFs. In \cite{p8}, the model requires an explicit VL representation. We adopt a resource oriented state representation. A simpler model considering latency remains an open topic of study.

\paragraph{Action representation}

In \cite{p1,rkhami2021learn,p4,p5,p2}, the problem is modeled by considering finite action spaces. The action in \cite{p1,p4,p5} is the index of the physical node in which to place a specific VNF of the slice. This representation of the action requires  breaking the process of placing one slice in a sequence of placement actions and has the advantage of reducing the size of the action space to the number of physical nodes. In \cite{rkhami2021learn}, the action is represented as a binary variable used to modify or accept the placement of a specific VNF of the slice previously computed by a heuristic. The action in \cite{p2} is either to mark a VNF to be placed on a physical node or to place a VNF on the previously marked physical node. The placement of the entire slice is then iteratively constructed and the algorithm stops when all the VNFs in the slice are placed or when a maximal number of iterations is reached. In \cite{p3,p8,quang2019deep}, infinite action spaces are adopted, i.e., the actions are real numbers. In \cite{p3} the action is defined as an instruction to a scheduler to adjust the resources allocated to a VNF by a certain percentage whereas in \cite{p8}, the action is the placement price returned to the client. 

In \cite{quang2019deep}, the action is represented by two sets of weights: i) the placement priority of each VNF in the slice on each physical node and ii) the placement priority of each VL in the slice on each physical link. To reduce complexity, we adopt a finite action space represented as  in \cite{p1}.
\paragraph{Reward function}
Most of analyzed solutions adopt placement cost or revenue in their reward function \cite{p1, rkhami2021learn, p2,p4,p5,p8}. Cost and revenue are mainly calculated in terms of resource consumption. Some solutions adopt a reward associated to acceptance ratio \cite{p1,p4,quang2019deep}. A penalty for SLA violations is only considered by \cite{p3}. The most complete reward function is the one proposed by \cite{p1} combining acceptance ratio and placement cost with load balance. We leverage on \cite{p1} reward function criteria but propose a formulation that reduces bias during training.

\subsubsection{Modeling Aspects of DRL}
\paragraph{Use of convolutions}
Most of the analyzed papers employ a Convolutional Neural Network (CNN) to perform automatic feature extraction except the approach proposed in paper \cite{p4}, which uses regular Deep Neural Network (DNN). Classical CNNs are limited by the fact that they only work on Euclidean objects (e.g., images, grids). DRL algorithms for network slice placement built upon this technique have reduced real-life applicability because they can not work on unstructured networks. To overcome this issue, two of the analyzed DRL solutions applied a special type of CNN, called GCN \cite{Defferrard,kipf_gcn}, which generalizes CNNs to work on arbitrarily structured graphs. Paper \cite{p1} uses GCN to automatically extract features from the physical network when solving a VNE problem. A type of GCN adapted to hetero-graphs called Relational GCN is used by \cite{rkhami2021learn} to automatically learn to improve the quality of an initial placement computed by a heuristic. We also use the power of GCN in our proposal.
\paragraph{Training algorithms}
All the analyzed solutions use different types of Policy Gradient (PG) training algorithms, except for the solution in paper \cite{p2} which uses the well-known value-based algorithm ``Deep Q-learning". We rely on the Asynchronous Advantage Actor Critic (A3C) approach introduced by \cite{a3c} and also used by \cite{p1} due to its robustness and improved performance.

\subsection{On Hybrid ML Approaches for Slice Placement \label{sec:hybrid}}

Although ML-based techniques, such as DRL, have been shown to be robust when applied to solving various problems, these approaches also exhibit some drawbacks as they: i) are difficult to run due to a variety of hyper parameters to control; ii) have convergence times  difficult to control  since they act as a self-controlled black box;  and iii) take too much time to start finding good solutions because their performance depends on the exploration of a huge number of states and actions. These make the application of pure DRL approaches risky in real online optimization scenarios. Researchers then focused on developing hybrid methods that combine ML-based optimization with safer but less scalable techniques such as heuristics. The concept of Heuristically Accelerated Reinforcement Learning (HARL) is introduced in \cite{bianchi2008accelerating} as a way to solve RL problems with the explicit use of a heuristic function to influence the learning agent's choice of actions. 

As shown by \cite{bianchi2012heuristically} HA- versions of well-known RL algorithms such as Q-learning, SARSA($\lambda$) and TD($\lambda$) have lower convergence times than those of their classical versions when applied to a variety of RL problems. Although HARL has shown relevant results in different fields of applications such as video streaming \cite{supancic2017tracking}, multi-agent systems \cite{bianchi2013heuristically} and robotics \cite{bianchi2018heuristically}, to the best of our knowledge, it has only been used by \cite{morozs2015heuristically} in the networking domain, to autonomous spectrum management. We build on HARL to propose HA-DRL and we apply it in the present paper to slice placement. To the best of our knowledge, this is the first work on HA-DRL. We rely on the formulation of heuristic function proposed in \cite{bianchi2008accelerating}. We adapt it to DRL and combine it with an efficient placement heuristic we proposed in \cite{cnsm_2020} to strengthen and accelerate the DRL learning process improving performance and safety of state-of-the-art DRL placement approaches \cite{p1}. 

Other hybrid approaches for placement optimization combining DRL and heuristics were proposed recently. In \cite{quang2019deep} a HFA-DDPG algorithm combining Deep Deterministic Policy Gradient (DDPG) with a Heuristic Fitting Algorithm (HFA) to solve the VNF-FGE problem is proposed. Evaluation results show that HFA-DDPG converges much faster than DDPG. However, the approach reveals to be more complex than needed due to the infinite action space formulation that adds an unnecessary complexity reducing the applicability of the algorithm. Our approach, with a bounded action space, reduces the complexity and seems more appropriated for large-scale scenarios. In \cite{rkhami2021learn}, the  REINFORCE algorithm is used to learn and to improve a VNE solution previously computed by a heuristic. In spite of  the significant improvement of the initial heuristic solution obtained by this approach, its effectiveness is highly dependent on the quality of the initial solution provided by the heuristic. In addition, both algorithms proposed in \cite{quang2019deep} and \cite{rkhami2021learn} are based on a DRL agent that relies heavily on the heuristic algorithm to compute the placement decisions. In our case, we only use the heuristic to improve the DRL exploration process. As a result, our DRL agent can scale and be used after training even without the support of the heuristic.

\section{Network Slice Placement Optimization Problem Modeling \label{sec:network_model}}

We present, in this section, the two main elements of  the Network Slice Placement Optimization problem: the Physical Substrate Network, in Section \ref{sec::psn_model}, and the Network Slice Placement Requests, in Section \ref{sec:nspr_model}. 


\subsection{Physical Substrate Network Modeling \label{sec::psn_model}}

\begin{figure}[hbtp]
\centering
\includegraphics[width=\linewidth]{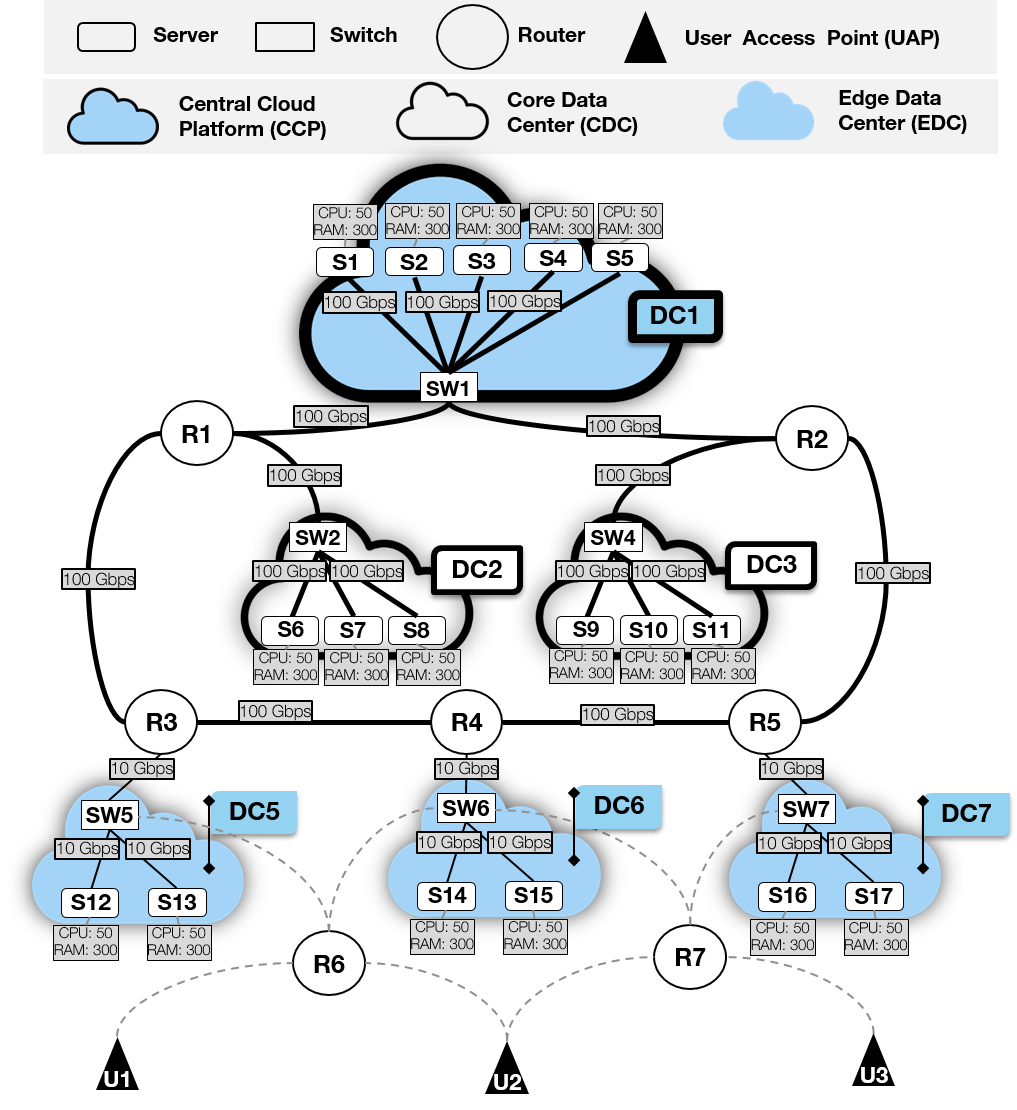}
\caption{Physical Substrate Network.}
\label{fig:sn_model}
\end{figure}

The Physical Substrate Network (for short, PSN) is composed of  the infrastructure resources. These ones are  IT resources (CPU, RAM, disk, etc.) needed for supporting  the VNFs of network slices  along with the transport network, in particular VLs for interconnecting VNFs of slices. As illustrated in Figure~\ref{fig:sn_model}, the PSN  is divided into three components: the Virtualized Infrastructure (VI) corresponding to IT resources, the Access Network (AN), and the Transport Network (TN). 

\subsubsection{The Virtualized Infrastructure (VI)} This component is the set of Data Centers (DCs) interconnected by network elements (switches and routers) and either distributed in Points of Presence (PoP) or centralized (e.g., in a big cloud platform). They offer IT resources to  run VNFs.

Following the reference architecture presented in \cite{slim2018close} and describing the architecture of a network operator, we define three types of DCs with different capacities: Edge Data Centers (EDCs) as local DCs with small resources capacities, Core Data Centers (CDCs) as regional DCs with medium resource capacities, and Central Cloud Platforms (CCPs) as national DCs with big resource capacities.

\subsubsection{The Access Network (AN)} This set represents User Access Points (UAPs) (Wi-Fi APs, antennas of cellular networks, etc.) and Access Links. Users access  slices  via one UAP, which may change during the life time of a communication by a  user (e.g., because of mobility).

\subsubsection{The Transport Network (TN)} This is the  set of routers and transmission links needed to interconnect the different DCs and the UAPs.

The complete PSN is modeled as a weighted undirected graph $G_s = (N, L)$ where $N$ is the set of physical nodes in the PSN, and $L \subset \{(a, b) \in N \times N \wedge a\neq b\}$ refers to a set of substrate links. Each node has a type in the set $\{$UAP, router, switch, server$\}$. The available CPU and RAM capacities on each node are defined respectively as $cap^{cpu}_n \in \mathbb{R}$, $cap^{ram}_n \in \mathbb{R}$ for all $n \in N$. The available bandwidth on the links are defined as $cap^{bw}_{(a,b)} \in \mathbb{R}, \forall (a,b) \in L$.

\subsection{Network Slice Placement Requests Modeling \label{sec:nspr_model}}

We consider each slice as a finite number of VNFs to be placed and chained on the PSN. VNFs are batched and introduced in the network as Network Slice Placement Requests (NSPRs). The NSPRs are similarly represented as a weighted undirected graph  $G_v = (V, E)$  where $V$ is the set of VNFs in the NSPR, and $ E \subset \{(\bar{a}, \bar{b}) \in V \times V \wedge \bar{a} \neq \bar{b}\}$ is a set of VLs. 

\begin{figure}[ht]
\centering
\includegraphics[width=\linewidth]{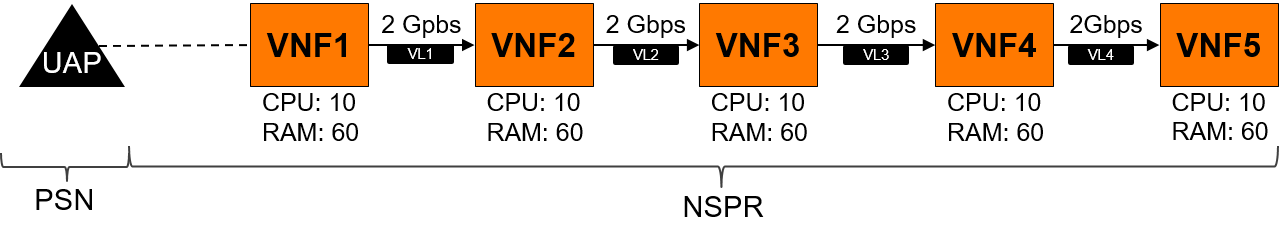}
\caption{Example of an NSPR.}
\label{fig:nspr_model}
\end{figure}

The CPU and RAM requirements of each VNF of a NSPR are defined as $req^{cpu}_{v} \in \mathbb{R}$ and $req^{ram}_{v} \in \mathbb{R}$ for all $v \in V$, respectively. The bandwidth required by each VL in a NSPR is given by $req_{(\bar{a},\bar{b})}^{bw} \in \mathbb{R}$  for all  $(\bar{a},\bar{b}) \in E$. An example of NPSR is represented in Figure~\ref{fig:nspr_model}.

\section{ Multi-objective Optimization Problem for Network Slice Placement \label{sec:problem_statement_and_formulation}}
In this section, we formulate the multi-objective optimization problem for network slice placement. The problem statement is given in Section \ref{sec:nsp_problem_statement} and  its ILP formulation in Section \ref{sec:problem_formulation}. The parameters related to PSN and NSPR are described in Tables \ref{tab::physical_substrate_network} and \ref{tab::nspr_parameters}, respectively.

\subsection{Network Slice Placement Optimization Problem Statement \label{sec:nsp_problem_statement}}


\begin{itemize}
    \item \textit{Given:} a NSPR graph $G_v = (V, E)$ and a PSN graph $G_s = (N, L)$,
    \item \textit{Find:} a mapping $G_v \to  \bar{G}_s =(\bar{N},\bar{L})$, $\bar{N} \subset N$, $\bar{L} \subset L$,
    \item\textit{Subject to:} the VNF CPU requirements $req^{cpu}_v, \forall v \in V$, the VNF RAM requirements $req^{ram}_v, \forall v \in V$, the VLs bandwidth requirements $req^{bw}_{(\bar{a},\bar{b})}, \forall (\bar{a},\bar{b}) \in E$, the server CPU available capacity $cap^{cpu}_s, \forall s \in S$, the server RAM available capacity $cap^{ram}_s, \forall s \in S$, the physical link bandwidth available capacity $cap^{bw}_{(a,b)}, \forall (a,b) \in L$.
    \item \textit{Objective: } maximize the network slice placement request acceptance ratio, minimize the total resource consumption and maximize load balancing.
\end{itemize}

\begin{table}[hbtp]
\caption{PSN parameters \label{tab::physical_substrate_network}}
\begin{tabular}{@{}cc@{}}
\toprule
\textit{\textbf{Parameter}}                          & \textit{\textbf{Description}}             \\ \midrule
$G_s = (N,L)$                                                  & PSN graph \\
$N$                                                  & Network nodes  \\
$S \subset N$                                        & Set of servers                            \\
$DC$                                                 & Set of data centers                       \\
$S_{dc} \subset S, \ \forall dc \in DC$              & Set of servers in data center $dc$        \\
$SW_{dc}, \ \forall dc \in DC$                       & Switch of data center $dc$  \\
$L = \{(a,b) \in N \times N \wedge a \neq b\}$ & Set of physical links                     \\
$cap^{bw}_{(a,b)} \in \mathbb{R}, \forall (a,b) \in L$ & Bandwidth capacity of  link $(a,b)$ \\
$cap^{cpu}_s \in \mathbb{R}, \forall s \in S$        & available CPU capacity on server $s$                \\
$M^{cpu}_s \in \mathbb{R}, \forall s \in S$        & maximum CPU capacity of server $s$                \\
$cap^{ram}_s \in \mathbb{R}, \forall s \in S$        & available RAM capacity on server $s$ \\
$M^{ram}_s \in \mathbb{R}, \forall s \in S$        & maximum RAM capacity of server $s$ \\
$M^{bw}_s \in \mathbb{R}, \forall s \in S$        & maximum outgoing bandwidth from $s$ \\ \bottomrule
\end{tabular}
\end{table}

\begin{table}[hbtp]
\centering
\caption{NSPR parameters \label{tab::nspr_parameters}}
\begin{tabular}{@{}cc@{}}
\toprule
\textit{\textbf{Parameter}}                                          & \textit{\textbf{Description}}                   \\ \midrule
$G_v = (V,E)$                                                                  & NSPR graph                         \\
$V$                                                                  & Set of VNFs of the NSPR                         \\
$E=\{(\bar{a},\bar{b}) \in N \times N \wedge \bar{a} \neq \bar{b}\}$ & Set of VLs of the NSPR                          \\
$req^{cpu}_{v} \in \mathbb{R}$                                         & CPU requirement of VNF $v$                      \\
$req^{ram}_{v} \in \mathbb{R}$                                         & RAM requirement of VNF $v$                      \\
$req_{(\bar{a},\bar{b})}^{bw} \in \mathbb{R}$                          & Bandwidth requirement of VL $ (\bar{a},\bar{b})$\\ \bottomrule
\end{tabular}
\end{table}

\subsection{Problem Formulation \label{sec:problem_formulation}}

To formulate the optimization problem, we introduce the decision  variables and we identify the constraints, which have to be satisfied by the placement algorithm.

\subsubsection{Decision Variables}

We use the  two following binary decision variables:
\begin{itemize}
 \item $x^{v}_{s} \in \{0,1\}$ for ${v} \in V$ and $s \in S$ is equal to 1 if the VNF $v$ is placed onto server $s$ and 0 otherwise,
 \item $y^{(\bar{a},\bar{b})}_{(a,b)} \in \{0,1\}$ for ${(\bar{a},\bar{b})} \in E$ and $(a,b) \in L$ is equal to 1 if the virtual link $(\bar{a},\bar{b})$ is mapped onto physical link $(a,b)$ and 0 otherwise.
\end{itemize}

\subsubsection{Problem Constraints}

\paragraph{VNF placement}
The following constraint ensures that: i) all VNFs of the NSPR must be placed; and ii) each VNF must be placed in only one server:
\begin{equation}
 \forall v \in V, \quad \sum_{s \in S} x^{v}_{s} = 1 .\label{c:node_mapping}   
\end{equation}

\paragraph{Network resource capacities constraints}
Constraints (\ref{c:cpu_requirements}) and (\ref{c:ram_requirements}) below ensure that the resource  capacities  of each server (for CPU and RAM, respectively) are not exceeded; the subsequent constraint~(\ref{c:bw_requirements}) guarantees that the bandwidth capacity of each physical link is not exceeded:
 \begin{align}
 \forall s \in S, & \quad \sum_{v \in V} d^{cpu}_{v}x^{v}_{s} \leq cap^{cpu}_s, && \label{c:cpu_requirements}\\
 \forall s \in S, & \quad \sum_{v \in V}d^{ram}_{v}x^{v}_{s} \leq cap^{ram}_s , \label{c:ram_requirements}\\
 \forall (a,b) \in L, & \quad \sum_{(\bar{a},\bar{b}) \in E}d^{bw}_{(\bar{a},\bar{b})}y^{(\bar{a},\bar{b})}_{(a,b)} \leq cap^{bw}_{(a,b)} . \label{c:bw_requirements}
 \end{align}

\paragraph{Eligible physical path calculation}

We use flow conservation constraints \cite{taccari2016integer} formulated by  Eq.~(\ref{c:link_mapping_1}), (\ref{c:link_mapping_2}), and (\ref{c:link_mapping_3}) to the definition of the eligible physical paths in which to map every VL $(\bar{a},\bar{b}) \in E$ as : for all $a \in S$ and $(\bar{a},\bar{b}) \in E$,

\begin{equation}
   \sum_{\substack{b \in N:\\ (a,b) \in L}} y^{(\bar{a},\bar{b})}_{(a,b)} - \sum_{\substack{b \in N: \\ (b,a) \in L}} y^{(\bar{a},\bar{b})}_{(b,a)}  = x^{\bar{b}}_{a} - x^{\bar{a}}_{a}, \label{c:link_mapping_1}  
\end{equation}
and
\begin{align}
& \forall a \in N \setminus S,    \forall (\bar{a},\bar{b}) \in E, \;    \sum_{\substack{b \in N:\\ (a,b) \in L}} y^{(\bar{a},\bar{b})}_{(a,b)} - \sum_{\substack{b \in N\\ (b,a) \in L}} y^{(\bar{a},\bar{b})}_{(b,a)} = 0,       \label{c:link_mapping_2} \\
 & \forall (\bar{a},\bar{b}) \in E,  \forall (a,b) \in L, \;     y^{(\bar{a},\bar{b})}_{(a,b)} + y^{(\bar{a},\bar{b})}_{(b,a)} \leq 1. \label{c:link_mapping_3} 
 \end{align}

The left hand sides of Eq.~(\ref{c:link_mapping_1}) and (\ref{c:link_mapping_2}) compute the difference between the activated links outgoing and incoming from/to each node $a \in N$. One computation is done for each VL $(\bar{a},\bar{b}) \in E$.  If $a \in S$, the right hand side of Eq.~(\ref{c:link_mapping_1}) ensures that the computed difference must be equal to $x^{\bar{b}}_{a} - x^{\bar{a}}_{a}$. That is: 0 or if server $a$ is used to place both VNFs $\bar{a}$ and $\bar{b}$ or if its not used to place neither of them; -1 if only the source VNF $\bar{a}$ is placed on $a$; 1 if only the destination VNF $\bar{b}$ is placed on $a$. If $a \in N \setminus S$, Constraint~(\ref{c:link_mapping_2}) ensures that this difference will always be  0. Constraint~(\ref{c:link_mapping_3}) imposes that each link must be used only in one direction when  mapping  a specific VL.

\subsubsection{Objective Function}
\label{sec:problem_formulation:obj}

We consider three objective functions: a) minimization of resource consumption; b) maximization of slice acceptance; and c) maximization of node load balance.

\paragraph{Minimization of resource consumption} The placement of all VNFs of an NSPR is mandatory otherwise the solution would violate Constraint~(\ref{c:node_mapping}). The optimization objective is then to minimize the bandwidth consumption given by 
\begin{equation}
  \min_{x,y}\sum_{(\bar{a},\bar{b}) \in E}\sum_{(a,b) \in L}y^{(\bar{a},\bar{b})}_{(a,b)}req^{bw}_{(\bar{a},\bar{b})}.  \label{obj_function}
\end{equation}
This objective can be seen as finding the shortest path to map each VL $(a,b) \in E$ of an NSPR. It is worth noting that Constraints~\eqref{c:link_mapping_1}, \eqref{c:link_mapping_2}, and \eqref{c:link_mapping_3}
 control the value of $y$ in such a way that $y^{(\bar{a},\bar{b})}_{(a,b)}=1$ for all links $(a,b) \in L$ included in the path used to map VL $(\bar{a},\bar{b}) \in E$. 

\paragraph{Maximization of slice requests acceptance} The maximization of slice request acceptance objective function is given by Equation (\ref{obj_function_2}). Auxiliary variable $z \in \{0,1\}$ representing whether the NSPR is accepted ($z = 1$) or not ($z = 0$)  is used in this case. 
\begin{equation}
  \max_{x,y} z . \label{obj_function_2}
\end{equation}

In this case, Constraint~(\ref{c:node_mapping}) is relaxed and additional Constraints~(\ref{c:request_mapping_1}) and (\ref{c:request_mapping_2}) below need to be inserted in the model:
\begin{align}
 \forall v \in V, \; & z \leq \sum_{s \in S} x^{v}_{s} , \label{c:request_mapping_1} \\
& z \geq \sum_{s \in S}\sum_{v \in V} x^{v}_{s} - |V-1| .\label{c:request_mapping_2}
\end{align}
Constraint (\ref{c:request_mapping_1}) ensures that $z = 0$ if there is a VNF $v \in V$ that cannot be placed (i.e., there is a $v \in V$ such that $\sum_{s \in S}x^{v}_{s} = 0$). Constraint (\ref{c:request_mapping_2}) ensures that $z=1$ if and only if all VNFs $ v \in V$ are placed (i.e., $\sum_{s \in S}\sum_{v \in V}x^{v}_{s} = |V|$).  

\paragraph{Maximization of node load balance}
Eq. (\ref{obj_function_3}) gives the maximization of node load balance objective function.
\begin{equation}
  \max_{x,y}\sum_{s\in S}\sum_{v\in V}x^{v}_{s}\left( \frac{cap^{cpu}_{s}}{M^{cpu}_{s}} + \frac{cap^{ram}_{s}}{M^{ram}_{s}} \right)  . \label{obj_function_3}
\end{equation}


\subsection{P2C Heuristic Principles\label{sec:P2C}}

We have proposed in \cite{cnsm_2020} a heuristic to solve the ILP problems introduced above. This heuristic is based on the P2C principle \cite{mitzenmacher2001power}, which states in the present context that considering 2 possible data centers chosen “randomly” instead of only 1 brings exponential improvement of the solution quality. 

The proposed heuristic is a greedy algorithm such that for each VNF $\bar{b} \in V$: i) randomly selects 2 candidate servers $s_1, s_2 \in S$; ii) evaluates the resource consumption when placing $\bar{b}$ in $s_1$ and $s_2$ and place $\bar{b}$ on the best server; iii) maps the VLs $(\bar{a},\bar{b}) \in E$ associated to $\bar{b}$. This heuristic contains the limitations of all heuristic approaches: the lack of flexibility due to manual feature design, the difficulties to handle multiple optimization criteria, and the sub-optimality of the provided solutions. But it also yields low execution time and good load balancing; it was shown in \cite{cnsm_2020} that the heuristic  outperforms two ILP based algorithms. We elaborate in the present paper on these proprieties of the heuristic to propose our HA-DRL approach described in the following sections. 

\section{DRL for Network Slice Placement Optimization  \label{sec:drl_proposal}}

In this section, we describe the DRL agent we propose to solve the ILP introduced in Section \ref{sec:problem_statement_and_formulation}. We give a general view of the DRL framework on Section \ref{sec:drl_framework} and describe the DRL framework's elements, the Policy, in Section \ref{sec::drl_policy}, the State, in Section \ref{sec::drl_state} and the Reward, in Section \ref{sec::drl_reward}.

\subsection{DRL Framework for Network Slice Placement\label{sec:drl_framework}}

\begin{figure}[hbtp] 
\centering
\includegraphics[width=\linewidth]{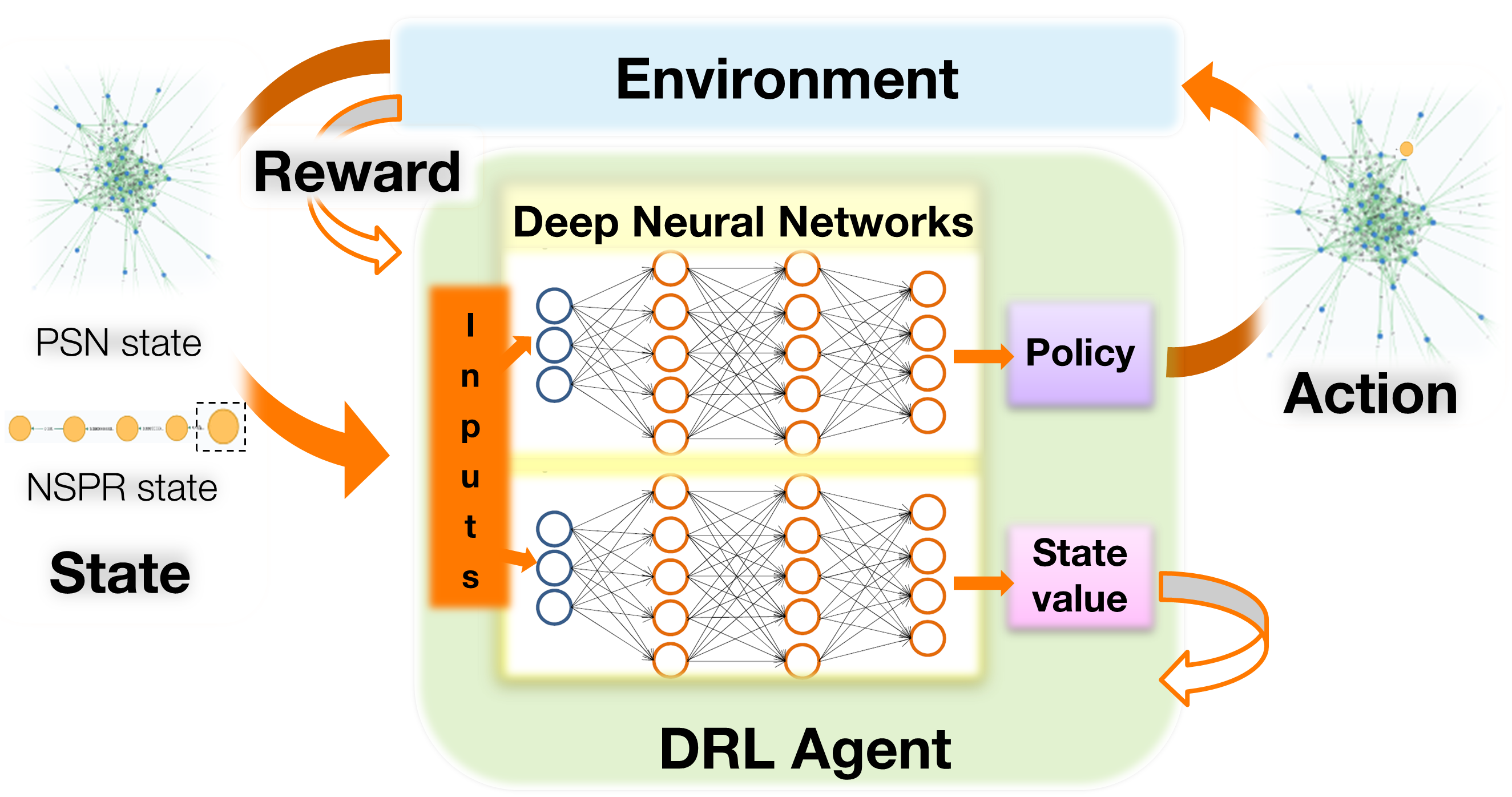}
\caption{DRL framework for Network Slice Placement Optimization} \label{fig::drl_framework_for_nsp}
\end{figure}

Figure~\ref{fig::drl_framework_for_nsp} presents the proposed DRL framework. The state contains the features of the PSN and NSPR to be placed. A valid action is, for a given NSPR graph $G_{v} = (V,E)$, a sub graph of the PSN graph $\bar{G_s} \subset \bar{G_s} = (N, L)$ to place the NSPR that does not violate the problem constraints described in Section \ref{sec:problem_formulation}. The reward evaluates how good is the computed action with respect to the optimization objectives described in Section \ref{sec:problem_formulation:obj}. DNNs are trained to: i) calculate optimal actions for each state (i.e., placements with maximal rewards), ii) calculate the State-value function used in the learning process.

\subsection{Policy Enforcement \label{sec::drl_policy}}


Let $\aset$ be the set of all possible actions that the DRL agent can take and $\sset$ the set of all states that it can visit. We adopt a sequential placement strategy in which, at each time step, we choose a node $n \in N$ where to place a specific VNF $v \in \{1,...,|V|\}$. The VNFs are placed in ascending order, which means that the placement starts with the VNF $v=1$ and ends for the  VNF $v = |V|$.

We break then the process of placing one NSPR graph $G_v = (V,E)$ in a sequence of $|V|$ actions, one for each $v \in V$, instead of considering the one shot placement of $G_v$. The latter strategy would require the definition of the action as a subgraph of the PSN graph $G_s = (N,L)$ what would imply $|\aset| = |SG|$, where $SG$ is the set of all sub graphs of $G_s$, that grows exponentially with size of $G_s$. Note that with the sequential placement strategy $\aset=N$, thus $|\aset| \ll |SG|$. At each time step $t$, the DRL agent focuses on  the placement of exactly one VNF $v \in V$ of the NSPR. 
Given a state $\sigma_t$, the DRL agent uses the policy to select an action $a_t \in \aset$ corresponding to the PSN node for placing VNF $v$. The policy probabilities are calculated using the Softmax distribution defined by
\begin{equation}
    \pi_{\theta}(a_{t} = a|\sigma_t) = \frac{e^{Z_{\theta}(\sigma_t,a),}}{\sum_{b \in N}e^{Z_{\theta}(\sigma_t,b)}},
    \label{eq::policy}
\end{equation}
where the  function $Z_{\theta}: \sset \times \aset \rightarrow \mathbb{R}$ maps each state and action to a real value. In our formulation this function is calculated by a DNN described in Section \ref{sec:dnn_structure}. The notation $\pi_{\theta}$ is used to indicate that policy depends on  $Z_{\theta}$. The control parameter $\theta$ represents the weights in the DNN.

\subsection{State Representation \label{sec::drl_state}}

The state contains a compact representation of the two main elements of our model.

\subsubsection{PSN State}

The PSN State is the real-time representation of the PSN status, which is given by four characteristics. Three of these characteristics are related to resources (see Table \ref{tab::physical_substrate_network} for the notation) and are defined by the following sets: $cap^{cpu} = \{cap^{cpu}_{n}: n \in N\}$, $cap^{ram} = \{cap^{ram}_{n}: n \in N\}$ and $cap^{bw} = \{cap^{bw}_{n} = \sum_{(n,b) \in L}cap^{bw}_{(n,b)}: n \in N\}$. We  consider in addition one characteristic related to placement in order to record the actions taken by the DRL agent during the placement of the current NSPR described by the  vector $\chi = \{\chi_{n} \in \{0,..,|V|\} : n \in N \}$, where $\chi_{n}$ corresponds to the number of VNFs of the current NSPR placed on node $n$. 

\subsubsection{NSPR State}

The NSPR State represents a view of the current placement. It is composed by four characteristics. Three resource requirement characteristics (see Table \ref{tab::nspr_parameters} for the notation) associated with the current VNF $v$ to be placed: $req^{cpu}_{v}$, $req^{ram}_{v}$ and $req^{bw}_{v} =  \sum_{(v,\bar{b}) \in E}req^{bw}_{(v,\bar{b})}$. We also consider the number $m_{v} = |V| - v + 1$ used to track the number of  VNFs still  to be placed at each time step. 



\subsection{Reward Function \label{sec::drl_reward}}

The proposed Reward function contains one reward value for each optimization objective introduced in Section \ref{sec:problem_formulation} as detailed in the following sections.

\subsubsection{Acceptance Reward}
Each Action may lead to a successful or unsuccessful placement depending on whether it respects the problem constraints for the candidate VNF $v$ and its associated VLs or not. We then define the Acceptance Reward value due to action $a_t$ as
\begin{equation}
     \delta^{a}_{t+1} = \left\{\begin{array}{lr}
    100, & \text{if $a_{t}$ is successful, }\\
    -100, & \text{otherwise. }
    \end{array}\right.
    \label{eq::acceptance_signal}
\end{equation}

\subsubsection{Resource Consumption Reward}
As above, we define the Resource Consumption Reward value for the placement of VNF $v$ via action $a_t$ as
\begin{equation}
     \delta^{c}_{t+1}= \left\{\begin{array}{lr}
    \frac{req^{bw}_{(v-1,v)}}{req^{bw}_{(v-1,v)}|P|} = \frac{1}{|P|}, & \text{if $|P|>0$, }\\
    1, & \text{otherwise. }
    \end{array}\right.
    \label{eq::resource_consumption_signal}
\end{equation}
where $P$ is the path used to place VL $(v-1,v)$. Note that a maximum  $\delta^{c}_{t+1} = 1$ is given when $|P|=0$, that is, when VNFs $v-1$ and $v$ are placed on the same server. 

\subsubsection{Load Balancing Reward}
Finally, we define the Load Balancing Reward value for the placement of VNF $v$ via $a_t$
\begin{equation}
    \delta^{b}_{t+1} = \frac{cap^{cpu}_{a_t}}{M^{cpu}_{a_{t}}} + \frac{cap^{ram}_{a_t}}{M^{ram}_{a_{t}}}.
    \label{eq::load_balancing_signal}
\end{equation}

\subsubsection{Global Reward}
On the basis of the three reward values introduced above, we define the global as 
\begin{equation}
     \small
     r_{t+1} = \left\{\begin{array}{lr}
     0, & \text{if $t < T$ and $a_{t}$ is  successful}\\
     \sum^{T}_{i=0} \delta^{a}_{i+1}\delta^{b}_{i+1}\delta^{c}_{i+1}, & \text{if $t = T$ and $a_{t}$ is successful}\\
     \delta^{a}_{t+1}, & \text{otherwise}
    \end{array}\right.
    \label{eq::reward_function}
\end{equation}
where $T$ is the number of iterations of a training episode and the quantities $\delta^{a}_{i+1}$, $\delta^{b}_{i+1}$, and $\delta^{c}_{i+1}$ are defined by Equations~\eqref{eq::acceptance_signal}, \eqref{eq::load_balancing_signal} and \eqref{eq::resource_consumption_signal}, respectively.  The notation $r_{t+1}$ is used  to emphasize that the reward obtained via action $a_t$ is provided by the environment after the action is performed, that is, at time step $t+1$.
In the present case, we consider $T \leq |V|$. Our reward function formulation is inspired by that  proposed in \cite{p1} but different. In \cite{p1}, partial rewards are admitted by considering that $r_{t+1}= \delta^{a}_{i+1}\delta^{b}_{i+1}\delta^{c}_{i+1}$. In practice this approach can lead to ineffective learning, since the agent can learn how to obtain good partial rewards without never reaching its true goal, that is, to place the entire NSPR. 

To address this issue, we propose to accumulate the intermediary rewards and return them to the agent only if all the VNFs of the NSPR are placed --second case of Eq. (\ref{eq::reward_function}). A 0 reward is given to the agent on the intermediary time steps, where a VNF is successfully placed (first case of Eq. (\ref{eq::reward_function})). If the agent takes an unsuccessful placement action, a negative reward is given (third case of Eq.(\ref{eq::reward_function})).

\section{Adaptation of DRL and Introduction of a Heuristic Function
\label{sec:aidedDRL}}

\subsection{Proposed Deep Reinforcement Learning Algorithm \label{sec::drl_learning}}

To learn the optimal policy, we use a single thread version of the A3C Algorithm introduced in \cite{a3c}. This algorithm has evidenced good results when applied to the VNE problem in \cite{p1}. A3C uses two DNNs that are trained in parallel: i) the Actor Network with the parameter $\theta$, which is used to generate the policy $\pi_{\theta}$ at each time step, and ii) the Critic Network with the parameter $\theta_{v}$ which generates an estimate $\nu^{\pi_{\theta}}_{\theta_{v}}(\sigma_t)$ for the  State-value function defined by $$\nu_{\pi}(t|\sigma)=\mathbb{E}_{\pi}\left[\sum^{T-t-1}_{k=0}\gamma^{k} r_{t+k+1} | \sigma_t = \sigma \right],$$
equal to the expected return when starting on state $\sigma$ and following policy $\pi$ thereafter with the discount parameter $\gamma$.

\subsubsection{DNNs Structure \label{sec:dnn_structure}}

\begin{figure}[hbtp] 
\centering
\includegraphics[width=\linewidth]{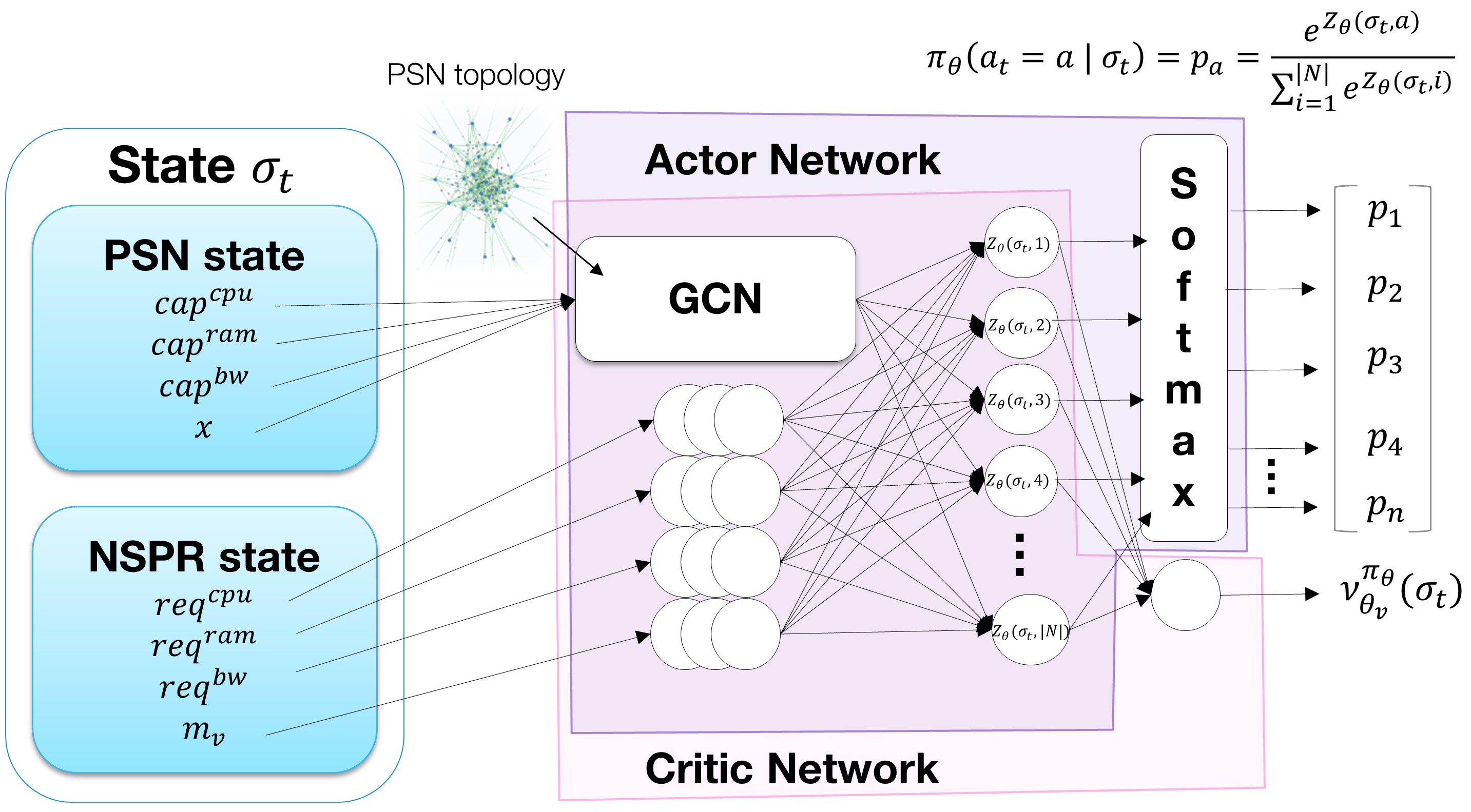}
\caption{Architecture of the proposed DRL algorithm} \label{fig::advantage_actor_critic_architecture}
\end{figure}

Both Actor and Critic Networks have identical structure except for their output layer as represented in Fig.~\ref{fig::advantage_actor_critic_architecture}. As in \cite{p1}, we use the GCN formulation proposed by \cite{kipf_gcn} to automatically extract advanced characteristics of the PSN. The characteristics produced by the GCN represent semantics of the PSN topology by encoding and accumulating characteristics of neighbour nodes in the PSN graph. The size of the neighbourhood is defined by the order-index parameter $K$. If $K$ is too large, the computation becomes expensive.

If $K$ is too small, the GCN will use information from a small subset of nodes and can become ineffective. The reader may refer to \cite{Defferrard,kipf_gcn} for more details. As in \cite{p1}, we consider in the following $K=3$ and perform automatic extraction of 60 characteristics per PSN node. 


The NSPR state characteristics are separately transmitted to a fully connected layer with 4 units. The characteristics extracted by both layers are combined into a single column vector of size $60|N| + 4$ and passed through a full connection layer with $|N|$ units. In the Critic Network, the outputs of this layer are forwarded to a single neuron, which is used to calculate the state-value function estimation $\nu^{\pi_{\theta}}_{\theta_{v}}(\sigma_t)$. In the Actor Network, the outputs of this layer represent the values of the function $Z_{\theta}$ introduced in Section \ref{sec::drl_policy}. These values are injected into a Softmax layer that transforms them into a Softmax distribution that corresponds to the policy $\pi_{\theta}$.

\subsubsection{DNNs Update} 

During the training phase, at each time step $t$, the  A3C algorithm uses the Actor Network to calculate the policy $\pi_{\theta}(.|\sigma_t)$. An action $a_t$ is sampled using the policy and performed on the environment. The Critic Network is used to calculate the state-value function approximation $\nu^{\pi_{\theta}}_{\theta_{v}}(\sigma_t)$. The agent receives then the reward $r_{t+1}$ and next state $\sigma_{t+1}$ from the environment and the placement process continues until a terminal state is reached, that is, until the Actor Network returns an unsuccessful action or until the current NSPR is completely placed. At the end of the training episode, the A3C algorithm updates parameters $\theta$ and $\theta_{v}$ adopting the  rules given in Table~\ref{updates}, derived from the PG Theorem \cite{a3c}.

\begin{table}[hbtp]
\caption{Updates of the control parameters $\theta$ and $\theta_v$.\label{updates}}
\hrule
\begin{align}
& J(\theta) = \sum^{T}_{t=t_{0}}\log\left(\pi_{\theta}(a_t|\sigma_t)\right) A^{\pi_{\theta}}(\sigma_t,a_t) \nonumber \\
& \delta(\theta) = \sum^{T}_{t=t_{0}} H(\pi_{\theta}(.|\sigma_t)) \nonumber \\
& \theta \leftarrow \theta + \frac{\alpha}{T - t_{0} +1} \left( \nabla_{\theta}J(\theta) + \phi\nabla_{\theta}\delta(\theta)\right)  \label{eq::update_rules_1}
\end{align}
\begin{align}
& J(\theta_v) = \sum^{T}_{t=t_{0}}(r_{t+1} + \nu^{\pi_{\theta}}_{\theta_{v}}(\sigma_{t+1}) - \nu^{\pi_{\theta}}_{\theta_{v}}(\sigma_{t}))^{2}    \nonumber \\
&  \theta_{v} \leftarrow \theta_{v} +\frac{\alpha'}{T - t_{0} +1} \nabla_{\theta_v}J(\theta_v)   \label{eq::update_rules_2}
\end{align}
\hrule
\end{table}

In Table~\ref{updates}, the term $A^{\pi_{\theta}}(\sigma_t,a_t)$ represents an estimate of the Advantage function, that indicates how better is the action $a_t$ when compared against  the “average action” from the corresponding policy under a certain state $\sigma_t$. By applying the Temporal Difference (TD) method \cite{sutton2018reinforcement}, $A^{\pi_{\theta}}(\sigma_t,a_t) = r_{t+1} + \nu^{\pi_{\theta}}_{\theta_{v}}(\sigma_{t+1}) - \nu^{\pi_{\theta}}_{\theta_{v}}(\sigma_{t})$.

The $J(\theta)$ function is the performance of the Actor Network. It is given by the Log-likelihood of the policy weighted by the Advantage Function. The $\delta(\theta)$ is the sum of the policy entropy $H(\pi_{\theta}(.|\sigma_t)$ used as a regularization term to discourage premature convergence. The function $J(\theta_{v})$ is the performance of the Critic Network. It is given by the squared Temporal Difference error of  $\nu^{\pi_{\theta}}_{\theta_{v}}(\sigma_{t+1})$. The gradients $\nabla_{\theta}$ and $\nabla_{\theta_{v}}$ are the vectors of partial derivatives w.r.t. $\theta$ and $\theta_v$, respectively. The hyper-parameters $\alpha$, $\alpha'$, and $\phi$ are the learning rates and heuristically fixed. $t_{0}$ is the first time step of the training episode and $T$ is the last one. Note that $t_{0}\leq T < t_{0} + |V|$.   

\subsection{Remarks on the Implementation of the DRL Algorithm}

All resource-related characteristics are normalized to be in the $[0,1]$ interval. This is done by dividing $cap^{j}$ and $req^{j}$, $j \in \{$cpu, ram,bw$\}$ by $\max_{n \in N}M^{j}_{n}$. With regard to the DNNs, we have implemented the Actor and Critic as two independent Neural Networks. Each neuron has a bias assigned. We have used the hyperbolic tangent (tanh) activation for non-output layers of the Actor Network and Rectified Linear Unit (ReLU) activation for all layers of the Critic Network. We have normalized positive global rewards to be in the $[0,10]$ interval. During the training phase, we have considered the policy as a Categorical distribution and used it to sample the actions randomly. In the validation tests, we have always selected  the action with higher probability, i.e., $a_t = \text{argmax}_{a \in N}\,\pi_{\theta}(a|\sigma_t)$. As  in \cite{p1}, we have taken  $\phi=0.5$. We performed hyper-parameter search to define $\alpha$ and $\alpha'$ values as described in Section \ref{sec:hps}.

\subsection{Introduction of a Heuristic}

\subsubsection{Motivation}
For the proposed DRL agent to learn the optimal policy $\pi^{*}_{\theta}$ it needs to perform actions over various states and receive the respective rewards to improve the policy iteratively. However, this is not a straightforward process since the convergence depends on many hyper parameters. The agent may get stuck in sub-optimal policies or have trouble to learn good estimations for the state-value function. Also, pure DRL approaches need to perform many actions and visit a huge number states to learn good policies. 

As a consequence, these approaches take much time to start providing good solutions. We propose then to reinforce and accelerate the DRL learning process using the heuristic described in Section \ref{sec:P2C}.

\subsubsection{Modified DRL algorithm}

\begin{figure}[hbtp] 
\centering
\includegraphics[width=\linewidth]{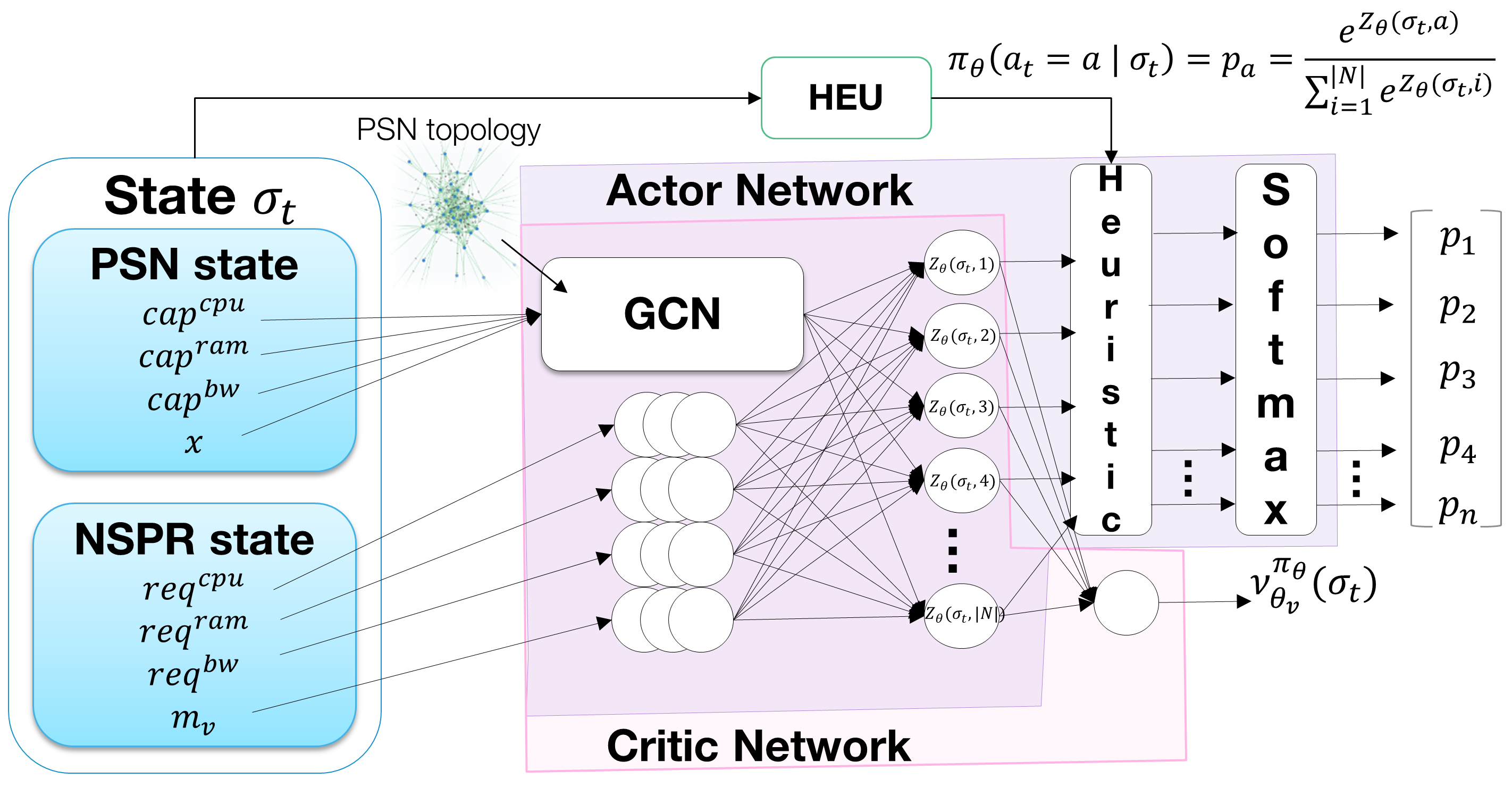}
\caption{Architecture of the proposed HA-DRL algorithm} \label{fig::ha_advantage_actor_critic_architecture}
\end{figure}

Fig.~\ref{fig::ha_advantage_actor_critic_architecture} presents the architecture of the proposed HA-DRL algorithm. We modify the structure of the Actor Network by introducing a new layer, namely the Heuristic layer that calculates an Heuristic Function $H: \sset \times \aset \rightarrow \mathbb{R}$ based on external information provided by the heuristic method described in Section~\ref{sec:P2C} and referred hereafter  as HEU.

As described in \cite{bianchi2012heuristically}, the Heuristic Function $H(\sigma_t,a_t)$ is a policy modifier that defines the importance of executing a certain action $a_t$ in state $\sigma_t$. In what follows, we adapt the $H(\sigma_t,a_t)$ formulation proposed in \cite{bianchi2012heuristically} to our DRL algorithm and describe how we use $H(\sigma_t,a_t)$ as a modifier of the Actor Network output to give more importance to the action selected by the HEU method. 

\subsubsection{Heuristic Function Formulation \label{sec:heuristic_function_formulation}} Let $Z_{\theta}$ be the function computed by the fully connected layer of the Actor Network that maps each state and action to a real value which is after converted by the Softmax layer into the selection probability of the respective action (see Section \ref{sec::drl_policy}). Let $\bar{a}_{t} = \text{argmax}_{a \in \aset}\,Z_{\theta}(\sigma_t,a)$ be the action with the highest $Z_{\theta}$ value for state $\sigma_{t}$. Let $a^{*}_{t}=HEU(\sigma_t)$ be the action derived by the HEU method at time step $t$ and the preferred action to be chosen. $H(\sigma_t,a^{*}_t)$ is shaped to allow the value of $Z_{\theta}(\sigma_t,a^{*}_t)$ to become closer to the value of $Z_{\theta}(\sigma_t,\bar{a}_t)$. The aim is to turn $a^{*}_t$ into one of the likeliest actions to be chosen by the policy.

The Heuristic Function is then formulated as
\begin{multline}
     \small
     H(\sigma_t,a_t) =   \left\{\begin{array}{lr}
     Z_{\theta}(\sigma_t,\bar{a}_{t}) -  Z_{\theta}(\sigma_t,a_t) + \eta, & \text{if $a_{t}=a^{*}_{t}$}\\
     0, & \text{otherwise}
    \end{array}\right.
    \label{eq::heuristic_function}
\end{multline}
where $\eta$ parameter is a small real number. During the training process the Heuristic layer calculates $H(\sigma_t,.)$ and updates the $Z_{\theta}(\sigma_t,.)$ values by using the following equation:
\begin{equation}
    Z_{\theta}(\sigma_t,.) = Z_{\theta}(\sigma_t,.) + \xi H(\sigma_t,.)^{\beta} \label{eq:z_update} 
\end{equation}
The Softmax layer then computes the policy using the modified $Z_{\theta}$. Note the action returned by $a^{*}_{t}$ will have a higher probability to be chosen. The $\xi$ and $\beta$ are parameters used to control how much HEU influence the policy.





\section{Implementation and  Evaluation Results \label{sec:evaluation}}

In this section, we present the implementation and experiments we conducted to evaluate the proposed HA-DRL approach.

\subsection{Implementation Details \& Simulator Settings}

\subsubsection{Experimental setting} We developed a simulator in Python containing: i) the elements of the Network Slice Placement Optimization problem (i.e., PSN and NSPR); ii) the DRL and HA-DRL algorithms. We used the PyTorch framework to implement the DNNs. We consider an implementation of the HEU algorithm \cite{cnsm_2020} in Julia as a benchmark in the performance evaluations. Experiments were run in a 2x6 cores @2.95Ghz 96GB machine.

\subsubsection{Physical Substrate Network Settings} \label{sec::substrate_network_settings}
We consider a PSN that could reflect the infrastructure of an operator such as Orange \cite{farah2}. In this network, three types of DCs are introduced as in Section~\ref{sec:network_model}. Each CDC is connected to three EDCs which are 100 km apart. CDCs are interconnected and connected to a CCP that is 300 km away.  The Tables \ref{tab:SN_DC_info} and \ref{tab:SN_inter_DC_info} summarize the DCs and transport links properties. The CPU and RAM capacities of each server are 50 and 300 units, respectively.

\begin{table}[hbtp]
\centering
\caption{Data centers description}
\begin{tabular}{@{}cccc@{}}
\toprule
\textbf{\begin{tabular}[c]{@{}c@{}}Data \\ center \\ type\end{tabular}} &
  \textbf{\begin{tabular}[c]{@{}c@{}}Number of \\ data \\ centers\end{tabular}} &
  \textbf{\begin{tabular}[c]{@{}c@{}}Number of \\ servers \\ per data center\end{tabular}} &
  \textbf{\begin{tabular}[c]{@{}c@{}}Intra data \\ center links \\ bandwith capcity\end{tabular}} \\ \midrule
CCP & 1  & 16 & 100 Gbps \\
CDC & 5  & 10 & 100 Gbps \\
EDC & 15 & 4  & 10 Gbps  \\ \bottomrule
\end{tabular}
\label{tab:SN_DC_info}
\end{table}

\begin{table}[hbtp]
\centering
\caption{Transport links capacities}
\begin{tabular}{@{}cccc@{}}
\toprule
\textbf{}    & \textbf{CCP} & \textbf{CDC} & \textbf{EDC} \\ \midrule
\textbf{CCP} & NA           & 100 Gbps     & 100 Gbps     \\
\textbf{CDC} & 100 Gbps     & 100 Gbps     & 100 Gbps     \\
\textbf{EDC} & 10 Gbps      & 10 Gbps      & 10 Gbps      \\ \bottomrule
\end{tabular}
\label{tab:SN_inter_DC_info}
\end{table}

\subsubsection{Network Slice Placement Requests Settings \label{sec::network_slice_placement_requests_settings}}

We consider NSPRs to have the Enhanced Mobile Broadband (eMBB) setting described in \cite{cnsm_2020}. Each NSPR is composed of 5 VNFs. Each VNF requires 25 units of CPU and 150 units of RAM. Each VL requires 2 Gbps of bandwidth.

\begin{figure*}[htp] 
\centering
\begin{subfloat}[Actor Network, $\rho=0.50$.]
 {\includegraphics[width=.23\linewidth]{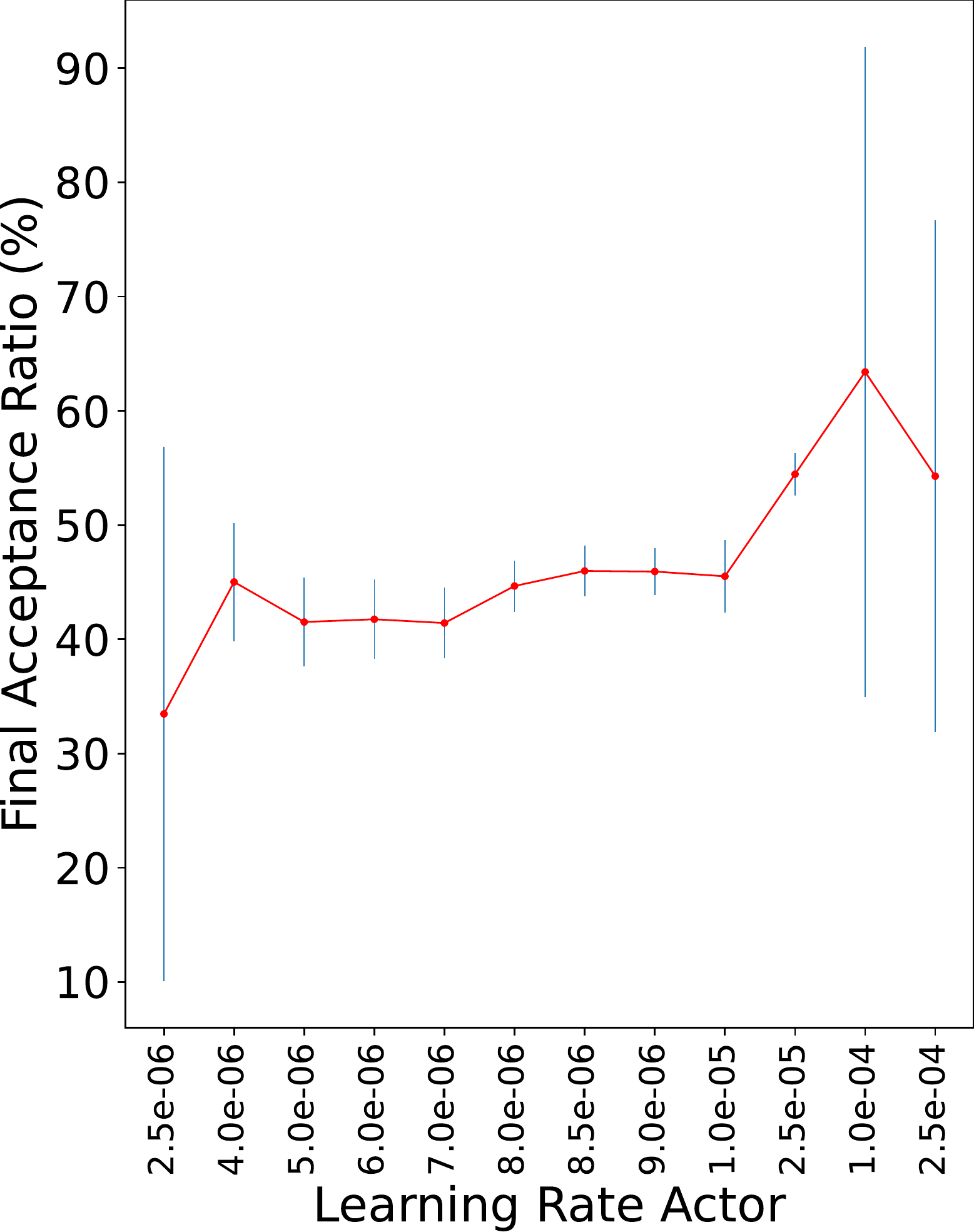}\label{fig:actor_ar_vs_lr_0.5}}
\end{subfloat}
\begin{subfloat}[Critic Network, $\rho=0.50$.]
 {\includegraphics[width=.23\linewidth]{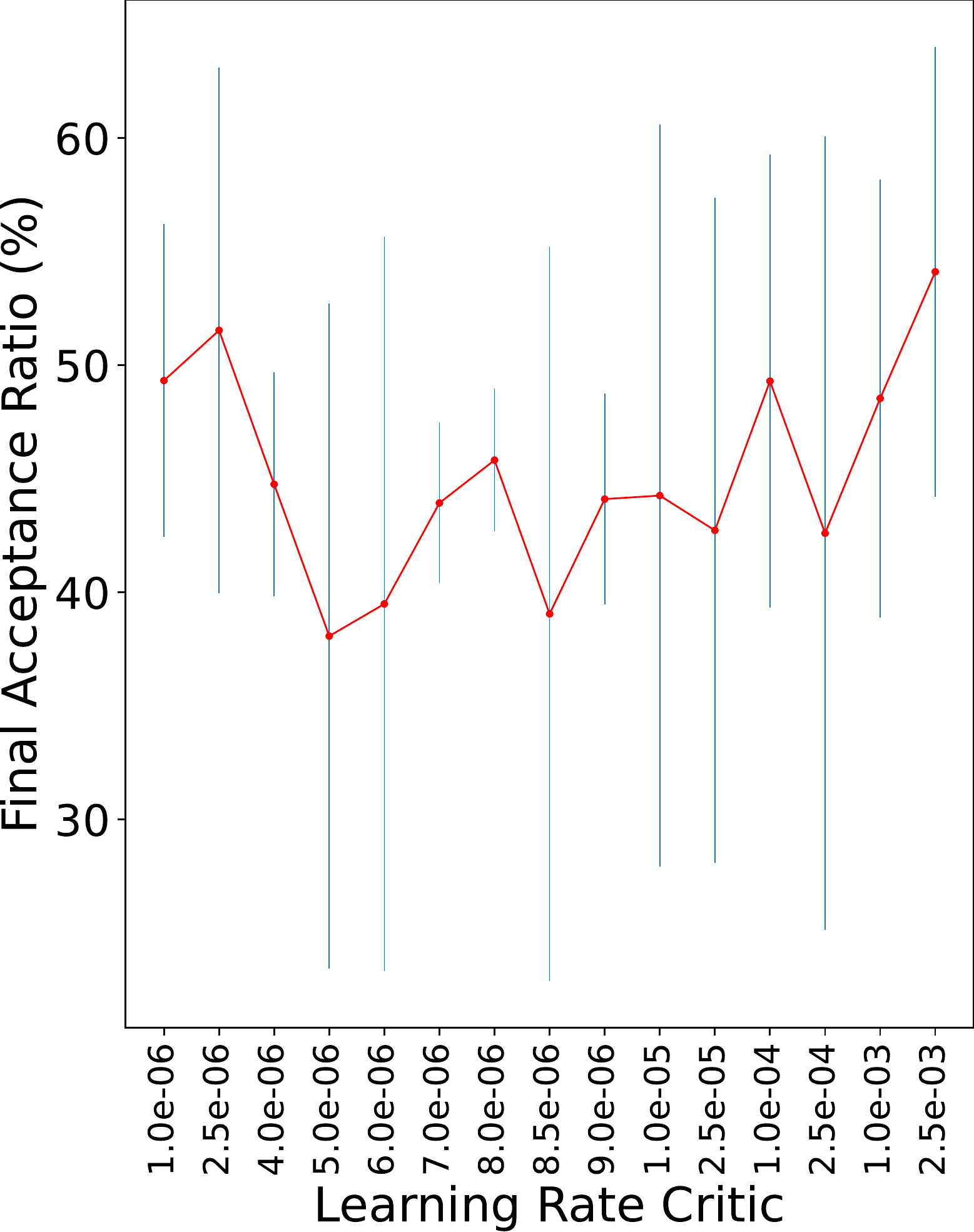} \label{fig:critic_ar_vs_lr_0.5}}
\end{subfloat}
\begin{subfloat}[Actor Network, $\rho=1.0$.]
 {\includegraphics[width=.23\linewidth]{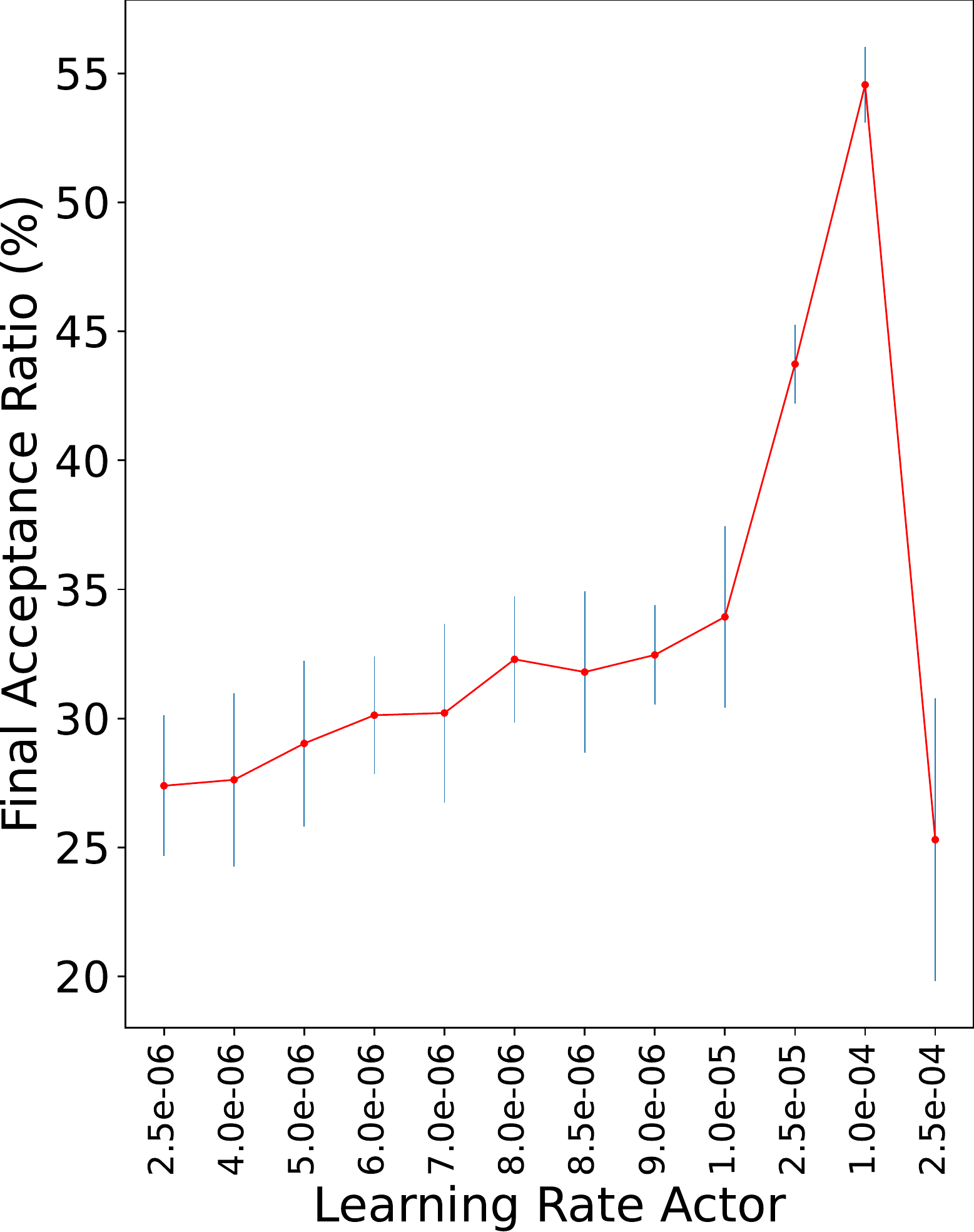}\label{fig:actor_ar_vs_lr_1.0}}
\end{subfloat}
\begin{subfloat}[Critic Network, $\rho=1.0$.]
 {\includegraphics[width=.23\linewidth]{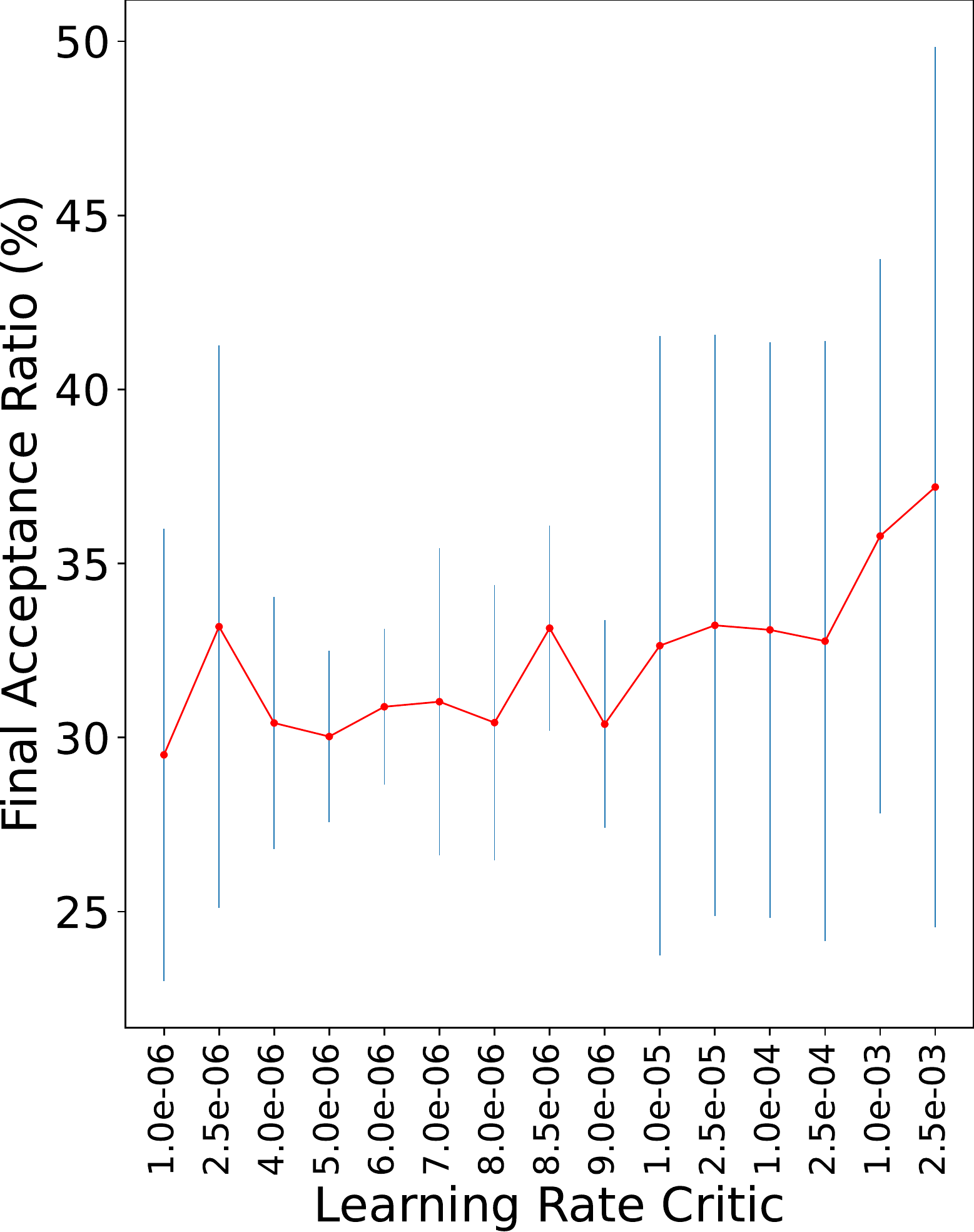}\label{fig:critic_ar_vs_lr_1.0}}
\end{subfloat}
\caption{Hyper-parameter search results.}
\label{fig:hps_results}
\end{figure*}

\subsection{Algorithms \& Experimental Setup }\label{sec:algorithms_tested}

\subsubsection{Training Process \& Hyper-parameters}
We consider a training process with maximum duration of 24 hours for the DRL and HA-DRL agents with learning rates for the Actor and Critic networks set to $\alpha = 10^{-4}$ and $\alpha' = 2.5. 10^{-3}$  (see Section \ref{sec:hps} about hyper-parameters tuning). We program four versions of HA-DRL agent (HA-DRL, $\beta = 0.1$; HA-DRL, $\beta = 0.5$; HA-DRL, $\beta = 1.0$; HA-DRL, $\beta = 2.0$), each with a different value for the $\beta$ parameter of the heuristic function formulation (see Section \ref{sec:heuristic_function_formulation}). We set the parameters $\xi = 1$ and $\eta = 0$.

\subsubsection{Heuristic baseline}
Since the performance of the HEU algorithm does not depend on learning, we consider the placement of 100,000 NSPRs with the HEU algorithm and use its performance in  the steady state as a benchmark.

\subsection{Network Load Calculation \label{sec::network_load_calculation}}
We use the formula proposed in \cite{farah2} to compute the NSPR arrival rates ($\lambda^{k}$) under the three network load conditions considered in the evaluation: underload ($\rho = 0.5$), normal load ($\rho = 0.8$ and $\rho = 0.9$), and critical load ($\rho = 1.0$). Network loads are calculated using CPU resources. In general, we set $1/\mu_{k} = 100$ time units for all $k \in \mathcal{K}$, where $\mathcal{K}$ is the number of  slice classes. For the resource $j$ with total capacity $C_j$, the load is $\rho_j = \frac{1}{C_j}\sum_{k=1}^{\mathcal{K}} \frac{\lambda^{k}}{\mu^{k}}A^{k}_{j}$, 
where $A^k_j$ is the number of resource units requested by an NSPR of class $k$.

\subsection{Evaluation Metrics \label{sec:ev_metrics}}
To characterize the performance of the placement algorithms, we consider 3 performance metrics: 
\begin{enumerate}
\item \textbf{Average execution time:} the average execution time in seconds required to place 1 NSPR. This metric is calculated based on 100 NSPR placements; 
\item \textbf{Acceptance Ratio per training phase}: the Acceptance Ratio obtained in each training phase, i.e., each part of the training process, corresponding to 1000 NSPR arrivals. It is calculated as follows: $\frac{\mathrm{\# accepted \; NSPRs}}{\mathrm{1000}}$. This metric is used to evaluate the convergence of the algorithms as it allows us to observe of the evolution of the agent's performance in time;
\item \textbf{Acceptance Ratio after training:} the Acceptance Ratio of the different tested algorithms after training computed after each arrival as follows: $\frac{\mathrm{\# accepted \; NSPRs}}{\mathrm{\# arrived \; NSPRs}}$. This metric is used in our validation test to compare the performance of the agents after training.
\end{enumerate}

\subsection{Hyper-parameters Tuning \label{sec:hps}}

To define the appropriate learning rates $\alpha$ and $\alpha'$ for training the Actor and Critic networks, respectively, we perform a classical hyper-parameter search procedure divided in 3 steps: i) we conduct several training experiments with different combinations of $\alpha$ and $\alpha'$ values, ii) we observe the final acceptance ratios, i.e., the acceptance ratios obtained in the last training phase of each experiment, and iii) we aggregate the final acceptance ratios by learning rate values. 

The DNNs are trained for 4 hours and we consider $\alpha$ and $\alpha'$ to be in the following intervals: $\alpha \in [0.00001,0.00025]$ and $\alpha' \in [0.000001,0.0025]$, where  upper bounds of the intervals are the learning rates used by \cite{p1}. Fig.~\ref{fig:hps_results} presents the average values and standard deviations of the final acceptance ratios aggregated by learning rate for the Actor and Critic networks considering two values for the network load parameter $\rho$: $\rho = 0.5$ (Fig.~\ref{fig:actor_ar_vs_lr_0.5} and \ref{fig:critic_ar_vs_lr_0.5}) and $\rho = 1.0$ (Fig.~\ref{fig:actor_ar_vs_lr_1.0} and \ref{fig:critic_ar_vs_lr_1.0}). As shown in Fig.~\ref{fig:actor_ar_vs_lr_0.5} and \ref{fig:actor_ar_vs_lr_1.0}, under both network load conditions, the best average of final acceptance ratios was achieved when considering a learning rate $\alpha = 10^{-4}$ for the Actor Network. This fact can be observed even more clearly when $\rho = 1.0$ since we observe a peak in the curve when $\alpha = 10^{-4}$ in addition to a smaller standard deviation as shown by Fig.~\ref{fig:actor_ar_vs_lr_1.0}. 

Fig.~\ref{fig:critic_ar_vs_lr_0.5} and \ref{fig:critic_ar_vs_lr_1.0} show that the best average of final acceptance rate was obtained when considering a learning rate $\alpha'= 2.5 \times 10^{-3} $ for the Critic Network. We also observe a higher standard deviation when aggregating the final acceptance ratios on the basis $\alpha'$ than on the basis of $\alpha$. The parameter $\alpha$ thus seems to have a stronger impact on the final acceptance ratios obtained than obtained for $\alpha'$.

\subsection{Acceptance Ratio Evaluation \label{sec:ar_evaluation}}

\begin{figure*}[t] 
\centering
\begin{subfloat}[$\rho=0.50$.]
 {\includegraphics[width=.23\linewidth]{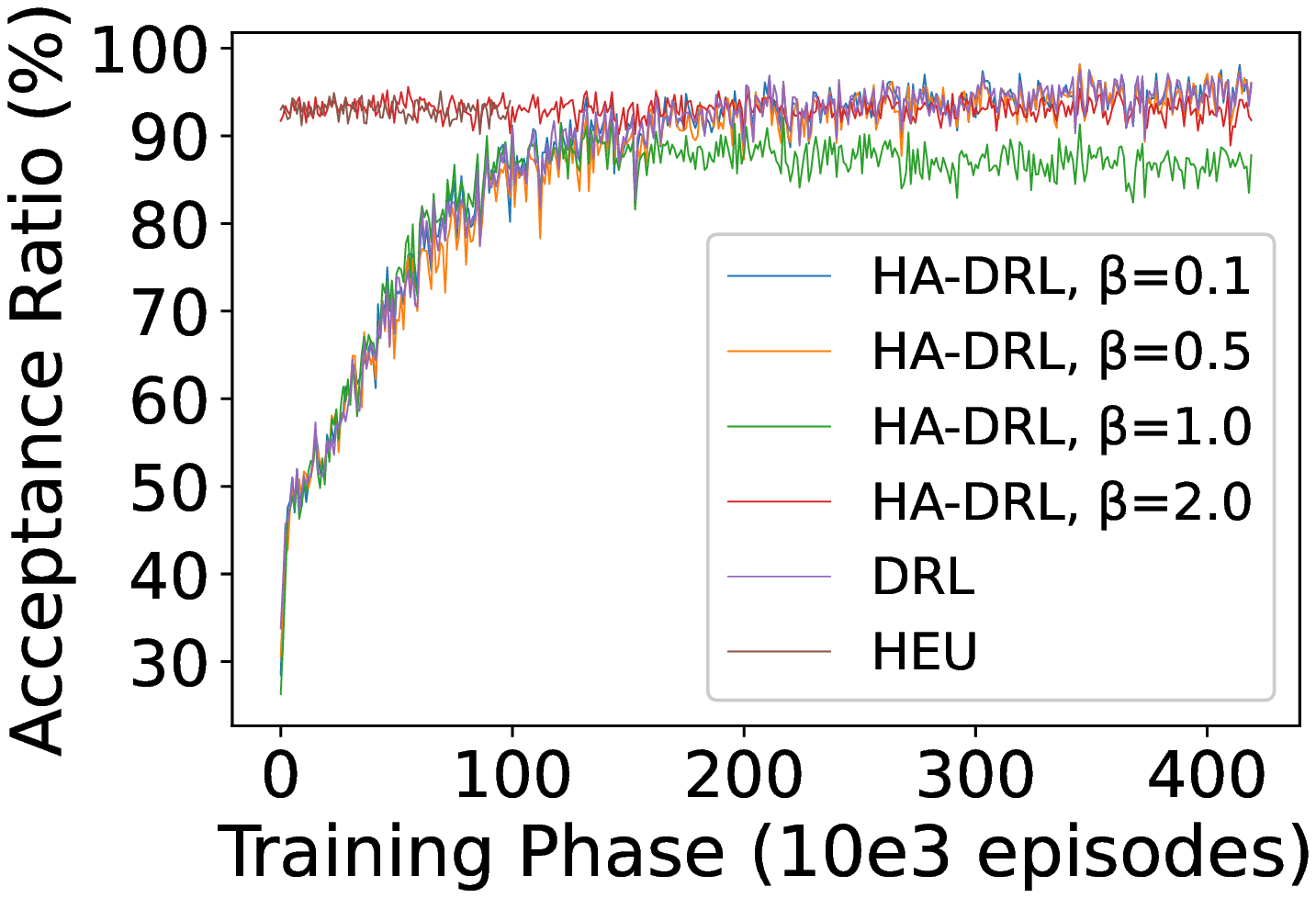}\label{fig:acc_ratio_0.5}}
\end{subfloat}
\begin{subfloat}[$\rho=0.80$.]
 {\includegraphics[width=.23\linewidth]{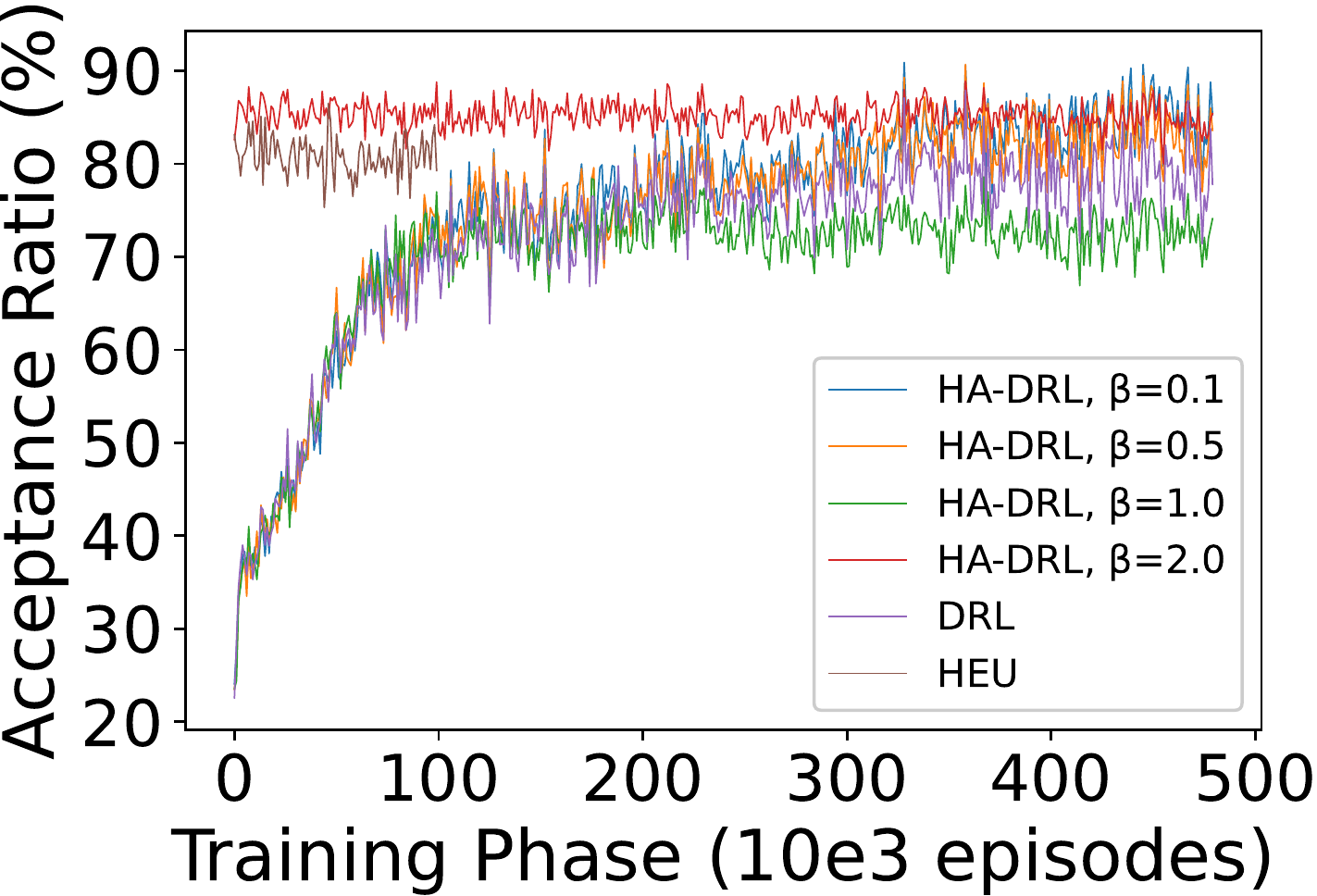}\label{fig:acc_ratio_0.8}}
\end{subfloat}
\begin{subfloat}[$\rho=0.90$.]
 {\includegraphics[width=.23\linewidth]{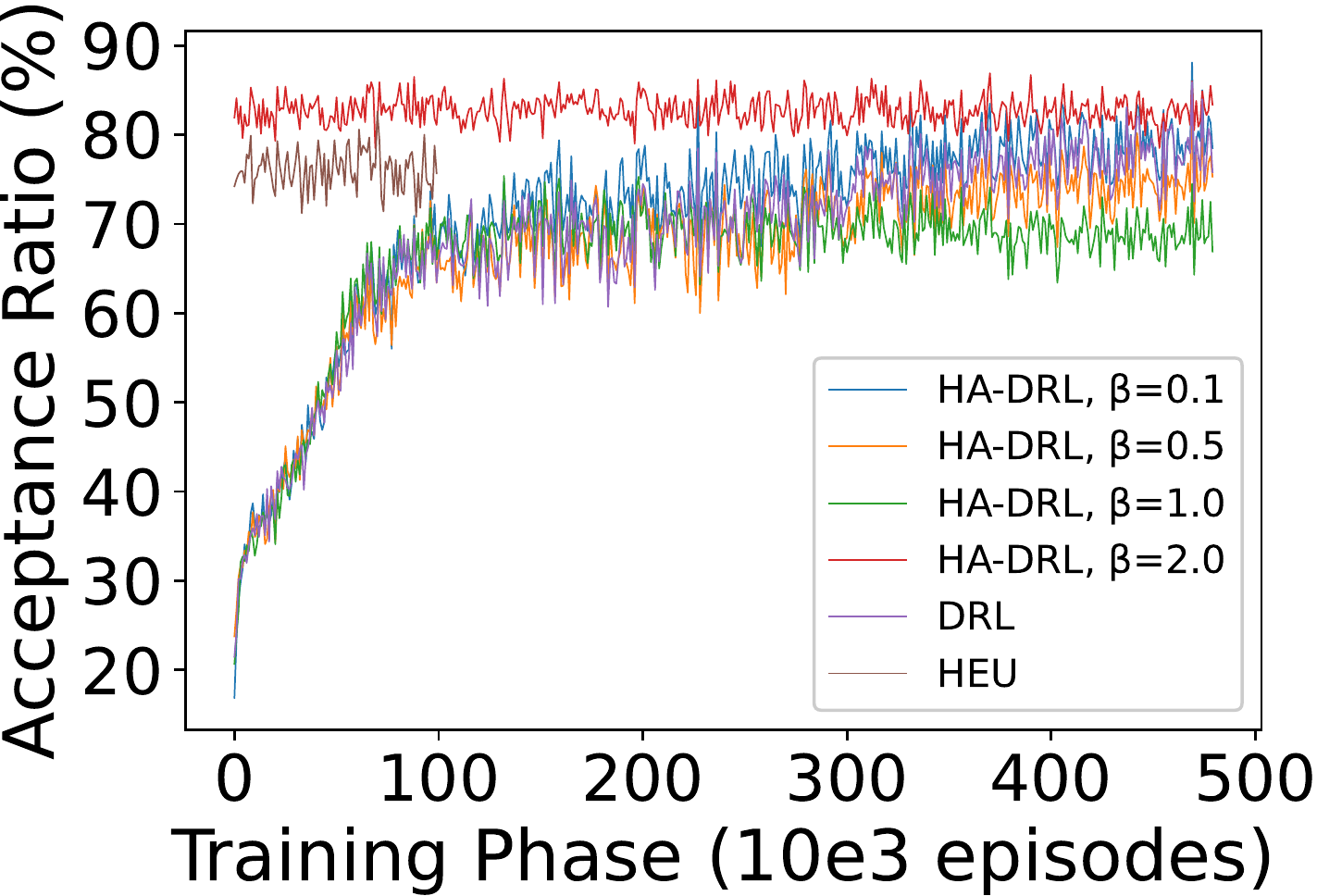}\label{fig:acc_ratio_0.9}}
\end{subfloat}
\begin{subfloat}[$\rho=1.0$.]
 {\includegraphics[width=.23\linewidth]{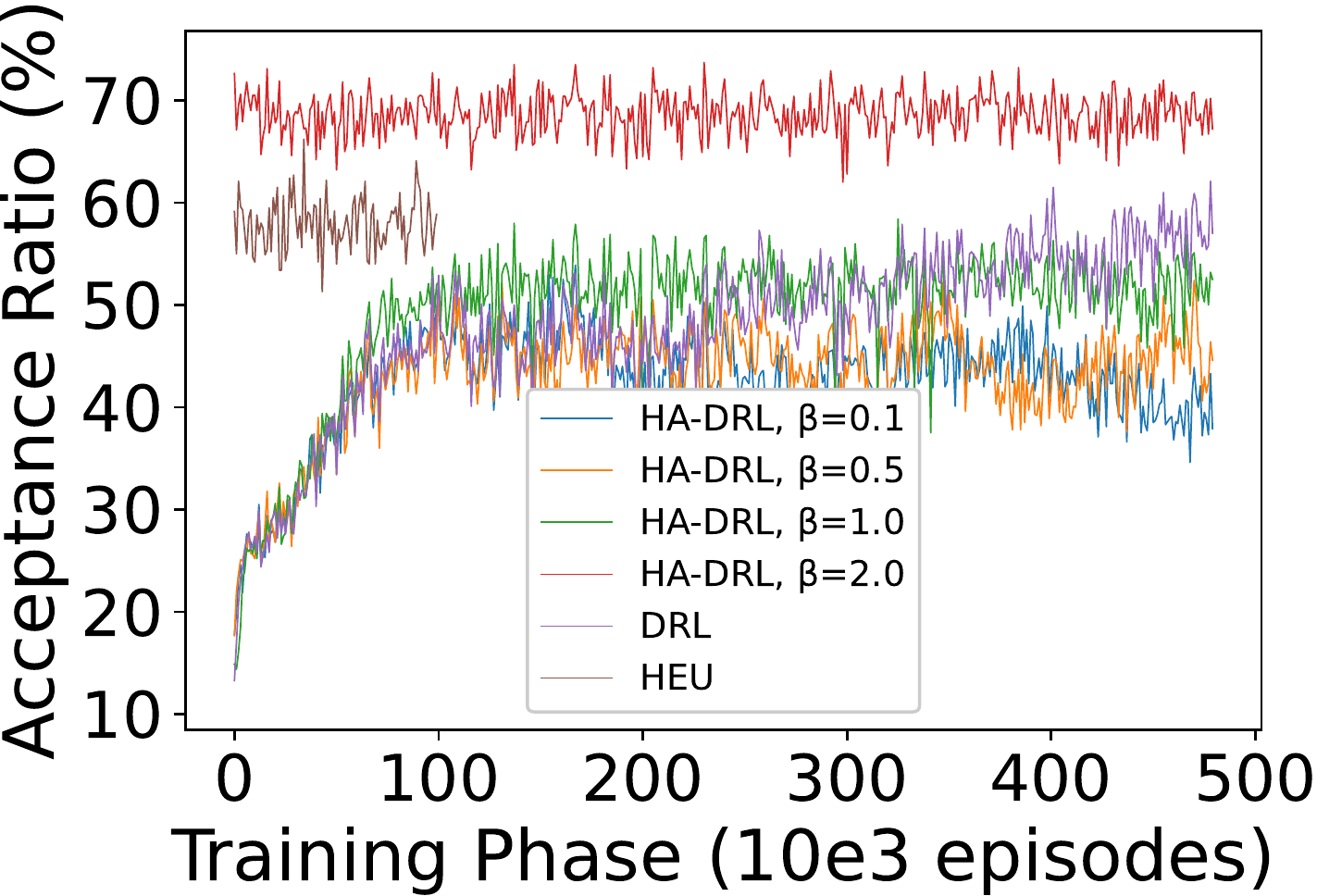}\label{fig:acc_ratio_1.0}}
\end{subfloat}
\caption{Acceptance ratio evaluation results.}
\label{fig:ar_vs_tr_phase}
\end{figure*}

Fig.~\ref{fig:ar_vs_tr_phase} and Tables \ref{tab:ars_0.5}-\ref{tab:ars_1.0} present the Acceptance Ratio per training phase obtained with the HA-DRL, DRL and HEU algorithms under different network loads. 

HA-DRL with $\beta = 2.0$ exhibits the most robust performance with convergence after a few training phases for all values of $\rho$. This happens because when setting $\beta = 2.0$ the Heuristic Function computed on the basis of the HEU algorithm has strong influence on the actions chosen by the agent. Since the HEU algorithm often indicates a good action, this strong influence of the heuristic function helps the algorithm to become stable more quickly. 

Fig.~\ref{fig:acc_ratio_0.5} and Table \ref{tab:ars_0.5} reveal that DRL and HA-DRL algorithms with $\beta \in \{0.1, 0.5,1.0\}$ converge within an interval of 200 to 300 training phases when $\rho=0.5$ and that all algorithms except HA-DRL with $\beta = 1.0$ have an Acceptance Ratio higher than 94\% in the last training phase. 

Fig.~\ref{fig:acc_ratio_0.8} and Table \ref{tab:ars_0.8} show that the performance of DRL and HA-DRL algorithms with $\beta \in \{0.1, 0.5,1.0\}$ stabilizes only after more than 400 training phases when $\rho =0.8$. They also show that when $\rho =0.8$, only HA-DRL with $\beta \in \{0.1, 0.5, 2.0\}$ have an Acceptance Ratio higher than 83\% in the last training phase. Algorithms HA-DRL with $\beta = 0.1$ and HA-DRL with $\beta = 2.0$ have similar performance and they are about 2\% better than HA-DRL for $\beta = 0.5$, 6\% better than HEU, 8\% better than DRL, and 11\% better than HA-DRL with $\beta = 1.0$. As shown in Fig.~\ref{fig:acc_ratio_0.9} and Table \ref{tab:ars_0.9} when $\rho =0.9$, only HA-DRL with $\beta = 2.0$ has an Acceptance Ratio higher than 83\% in the last training phase. HA-DRL with $\beta = 2.0$ is about 5\% better HA-DRL with $\beta = 0.1$,  8\% better than DRL, HEU and HA-DRL with $\beta = 0.5$, and  16\% better than HA-DRL with $\beta = 1.0$.

Fig.~\ref{fig:acc_ratio_0.9} and Table \ref{tab:ars_0.9} also reveal that the performance of algorithms DRL and HA-DRL with $\beta \in \{0.1, 0.5,1.0\}$ only stabilizes after more than 400 training phases when $\rho =0.9$. Fig.~\ref{fig:acc_ratio_1.0} and Table \ref{tab:ars_1.0} show that when $\rho =1.0$, HA-DRL with $\beta=2.0$ performs significantly better than the other algorithms at the end of the training. HA-DRL with $\beta=2.0$ accepts 67.2 \% of the NSPR arrivals in the last training phase, 8.34\% more than HEU, 10.21\% more than DRL,  14.7\% more than HA-DRL with  $\beta = 1.0$, 22.6\% more than HA-DRL with $\beta = 0.5$, 29.3\% more than HA-DRL with $\beta = 0.1$. Fig.~\ref{fig:acc_ratio_1.0} and Table \ref{tab:ars_1.0} also show that performance of algorithms DRL and HA-DRL with $\beta \in \{0.1, 0.5,1.0\}$ is still not stable at the end of the training process when $\rho =1.0$.

\begin{table}
\centering
\caption{Acceptance Ratio at different Training Phases, $\rho = 0.5.$}
\label{tab:ars_0.5}
\begin{tabularx}{\linewidth}{@{}cLLLLL@{}}
\toprule
\multirow{2}{*}{\textbf{Algorithm}} & \multicolumn{5}{c}{\textbf{Acceptance Ratio at different Training Phases (\%)}} \\ \cmidrule(l){2-6} 
                   &25&100&200&300&400    \\ \midrule
HADRL,$\beta$=0.1&57.90& 80.20&92.70&93.00&96.20\\
HADRL,$\beta$=0.5&58.50& 83.00&90.20&94.80&96.30\\
HADRL,$\beta$=1.0&58.80& 86.20&86.80&85.50&85.80\\
HADRL,$\beta$=2.0&93.50& 90.30&93.10&92.20&94.80\\
DRL                &57.10& 83.60&91.20&94.80& 95.70\\
HEU                &93.50& 94.00&94.00*&94.00 &94.00*\\ \bottomrule
\end{tabularx}
\end{table}

\begin{table}
\centering
\caption{Acceptance Ratio at different Training Phases, $\rho = 0.8.$}
\label{tab:ars_0.8}
\begin{tabularx}{\linewidth}{@{}cLLLLLL@{}}
\toprule
\multirow{2}{*}{\textbf{Algorithm}} & \multicolumn{6}{c}{\textbf{Acceptance Ratios at different Training Phases (\%)}} \\ \cmidrule(l){2-7} 
                                    &25& 100& 200&300& 400& 480\\ \midrule
HADRL,$\beta$=0.1&44.10&74.60&77.80&82.80&87.50&85.30\\
HADRL,$\beta$=0.5&46.30&75.00&77.40&80.90&86.90& 83.60\\
HADRL,$\beta$=1.0&46.00&77.00&74.90&72.40&74.80&74.10\\
HADRL,$\beta$=2.0&87.80&88.80&85.60&86.50&84.90&85.40\\
DRL&46.10&71.89&75.70&78.40&83.70& 77.80\\
HEU&79.20&79.27&79.27*&79.27*&79.27*&79.27*\\ \bottomrule
\end{tabularx}
\end{table}

\begin{table}
\caption{Acceptance Ratio at different Training Phases, $\rho = 0.9.$}
\label{tab:ars_0.9}
\begin{tabularx}{\linewidth}{@{}cLLLLLL@{}}
\toprule
\multirow{2}{*}{\textbf{Algorithm}} & \multicolumn{6}{c}{\textbf{Acceptance Ratios at different Training Phases (\%)}} \\ \cmidrule(l){2-7} 
                                    &25&100&200&300&400&480\\ \midrule
HADRL,$\beta$=0.1&42.40&66.20&75.00&73.00&81.70&78.50\\
HADRL,$\beta$=0.5&40.30&63.40&70.90&68.20&78.20&75.30\\
HADRL,$\beta$=1.0&42.69&66.00&70.90&68.40&71.10&66.90\\
HADRL,$\beta$=2.0&82.89& 81.10&84.10&81.00&82.80&83.36\\
DRL&41.60&63.40&72.10&71.10&80.60&75.80\\
HEU&73.90&75.68&75.68*&75.68*&75.68*&75.68*\\ \bottomrule
\end{tabularx}
\end{table}

\begin{table}
\caption{Acceptance Ratio at different Training Phases, $\rho = 1.0.$}
\label{tab:ars_1.0}
\begin{tabularx}{\linewidth}{@{}cLLLLLL@{}}
\toprule
\multirow{2}{*}{\textbf{Algorithm}} & \multicolumn{6}{c}{\textbf{Acceptance Ratios at different Training Phases (\%)}} \\ \cmidrule(l){2-7} 
                                    &25&100&200&300&400&480\\ \midrule
HADRL,$\beta$=0.1&30.10&49.70&45.30&45.0&41.00&37.90\\
HADRL,$\beta$=0.5&30.80&45.90&50.80&44.5&38.90&44.60\\
HADRL,$\beta$=1.0&27.40& 52.00&55.50&55.60&49.00&52.50\\
HADRL,$\beta$=2.0&67.60&67.60&70.80&69.60&66.10&67.2\\
DRL&29.30&49.10&46.60&50.10&53.40&56.99\\
HEU&60.70&58.86&58.86*&58.86*&58.86*&58.86*\\ \bottomrule
\end{tabularx}
\end{table}

\subsection{Execution Time Evaluation \label{sec:ex_time_evaluation}}

\begin{figure} 
\centering
\begin{subfloat}[]
 {\includegraphics[width=0.48\linewidth]{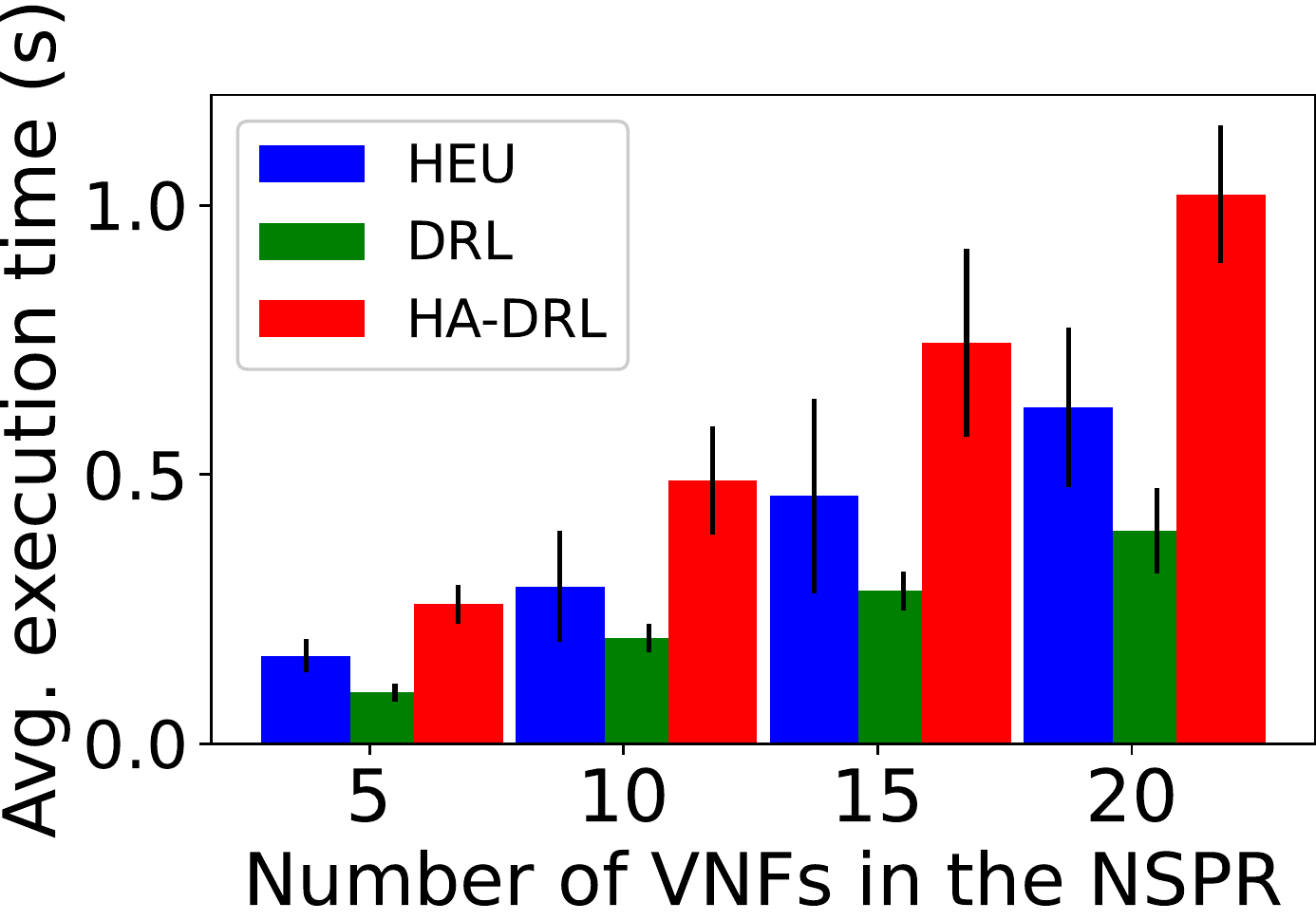}\label{fig:avg_exec_time_vs_nspr_size}}
\end{subfloat}
\begin{subfloat}[]
 {\includegraphics[width=0.48\linewidth]{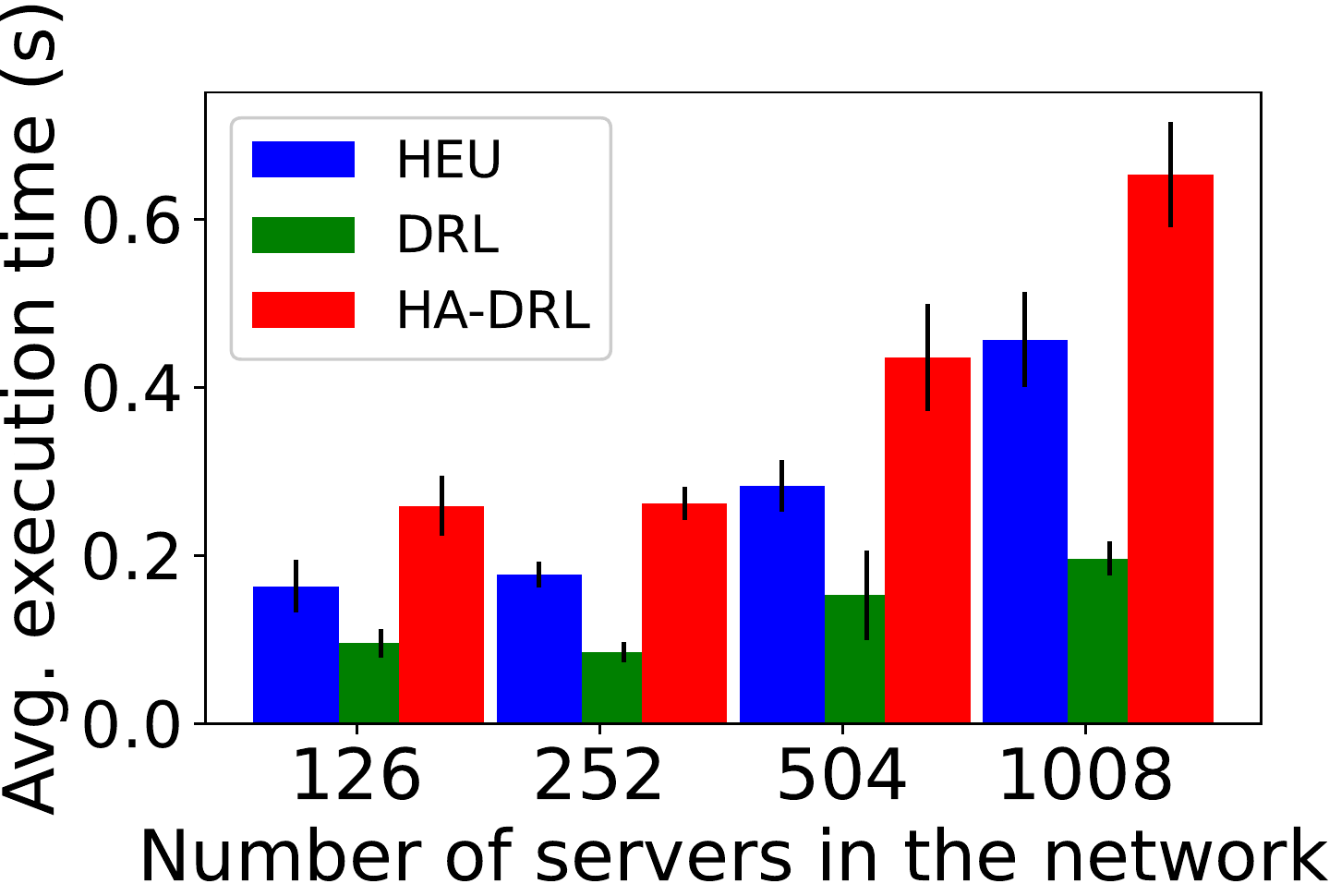}\label{fig:avg_exec_time_vs_nodes}}
\end{subfloat}
\caption{Average execution time evaluation.}
\label{fig:avg_exec_time}
\end{figure}

Fig.~\ref{fig:avg_exec_time_vs_nspr_size} and \ref{fig:avg_exec_time_vs_nodes} present the average execution time of the HEU, DRL and HA-DRL algorithms as a function of the number of VNFs in the NSPR and the number of servers in the PSN, respectively (see Section \ref{sec:ev_metrics} for details on the metric calculation). In both evaluations, we start by the PSN and NSPR settings described in Sections \ref{sec::substrate_network_settings} and \ref{sec::network_slice_placement_requests_settings}, respectively, and generate new settings by increasing either the number of VNFs per NSPR or the number of servers in the PSN. The evaluation results confirme our expectations by showing that the average execution times increase faster for heuristics than for a pure DRL approach. However, both HEU and DRL strategies have low execution times (less than 0.6s in the largest scenarios). The number of VNFs per NSPR has more impact on the average execution times of HEU and DRL algorithms than the number of servers on the PSN. The average execution time of HEU algorithm is more impacted than DRL by the number  of servers in the PSN. The HA-DRL algorithm depends on a sequential execution of DRL and HEU. Therefore, the average execution time of HA-DRL  is approximately to the sum of the execution times of HEU and DRL. Since DRL and HEU have small execution times, the average execution times of HA-DRL are also small (less than  1.0s for the largest NSPR setting and about 0.6s for the largest PSN setting).

\subsection{Validation Test \label{sec:val_tests}}

To validate the effectiveness of the different trained DRL agents, we perform a validation test. We consider the same PSN and NSPR settings described in Section \ref{sec:evaluation} and the arrival of 10,000 NSPRs to be placed and generating a network load $\rho = 0.8$. We run a simulation with each one of the trained DRL agents as well with the HEU algorithm and compare the obtained Acceptance Ratios at the end of the simulations (see description of the Acceptance Ratio after training metric on Section \ref{sec:ev_metrics}). 

\begin{figure}[t] 
\centering
\includegraphics[width=\linewidth]{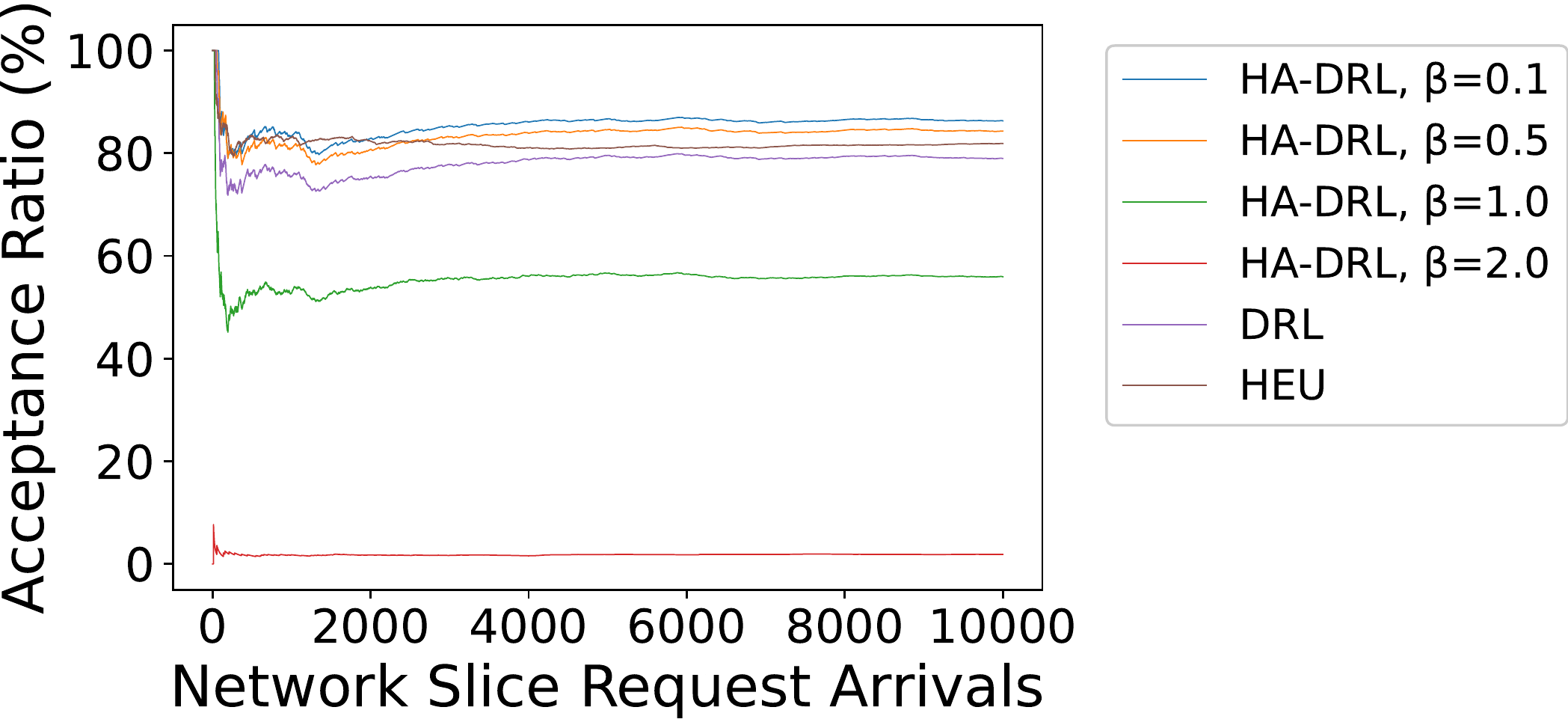}
\caption{Acceptance Ratio at validation test, $\rho=0.8$} \label{fig::validation_test_0.8}
\end{figure}

Fig.~\ref{fig::validation_test_0.8} and Table \ref{tab:validation_test} show that the HA-DRL agent  with $\beta=0.1$ is the one that better scales since it has the best Acceptance Ratio after training. HA-DRL agent with $\beta=2.0$ has the best Acceptance Ratio during training as described in Section \ref{sec:ar_evaluation} but is the one that scales the worst since it has poorest Acceptance Ratio after training performance. This is because with the use of a high $\beta$ value, HA-DRL suffers from a strong dependence on the HEU algorithm. This prevents HA-DRL with $\beta=2.0$ from being used  with the HEU algorithm deactivated. Note, however, that HA-DRL remains the best solution even when HEU is disabled after 24 hours of training since HA-DRL with $\beta=0.1$ has an Acceptance Ratio 7.34\% higher than the pure DRL and 4.42\% higher than the HEU algorithm.  

\begin{table}[ht]
\centering
\caption{Values of Acceptance Ratio after Training}
\label{tab:validation_test}
\begin{tabular}{@{}cc@{}}
\toprule
HADRL,$\beta=0.1$ & 86.31\% \\ \midrule
HADRL,$\beta=0.5$ & 84.31\% \\ \midrule
HADRL,$\beta=1.0$ & 55.95\% \\ \midrule
HADRL,$\beta=2.0$ & 1.89\%  \\ \midrule
DRL           & 78.97\% \\ \midrule
HEU           & 81.89\% \\ \bottomrule
\end{tabular}
\end{table}

\section{Conclusion\label{sec:conclusion}}

We have presented a Heuristically-assisted DRL (HA-DRL) approach to Network Slice Placement Optimization with 5 main contributions:  the proposed method i) enhances the scalability of existing ILP and heuristic approaches, ii) can cope with multiple optimization criteria, iii) combines DRL with GCN to automate feature extraction, iv)  strengthens and accelerates the DRL learning process using an efficient placement optimization heuristic, and  v)  supports multi-domain slice placement. 

Evaluation results show that the proposed HA-DRL approach yields good placement solutions in nearly real time, converges significantly faster than pure DRL approaches, and yields better performance in terms of acceptance ratio than state-of-the-art heuristics and pure DRL algorithms during and after the training phase. As a future work, we plan to explore a parallel computing implementation of HA-DRL to reduce its execution time and thus achieve the best performance in both Acceptance Ratio and Execution Time. We also plan to evaluate the HA-DRL in a online optimization scenario with fluctuating traffic demand to access the advantages of using HA-DRL in practice.

\section*{Acknowledgment}

This work has been performed in the framework of 5GPPP MON-B5G project (www.monb5g.eu). The experiments were conducted using Grid'5000, a large scale testbed by Inria and Sorbonne University (www.grid5000.fr).


\ifCLASSOPTIONcaptionsoff
  \newpage
\fi

\bibliographystyle{IEEEtran}
\bibliography{IEEEabrv,my_bib}



\begin{IEEEbiography}[{\includegraphics[width=1in,height=1.25in,clip,keepaspectratio]{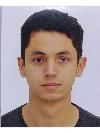}}]{José~Jurandir~Alves~Esteves} graduated from the University of Clermont Auvergne in 2017 and from the Federal University of Minas Gerais in 2019 obtaining two engineering degrees and a master's degree in computer science from the University of Clermont Auvergne. He is doing his PhD between Orange Labs and the Computer Science Laboratory of Paris 6 (LIP6), Sorbonne University. His research is related to automation and optimization models and algorithms for network slice orchestration.
\end{IEEEbiography}

\begin{IEEEbiography}[{\includegraphics[width=1in,height=1.5in,clip,keepaspectratio]{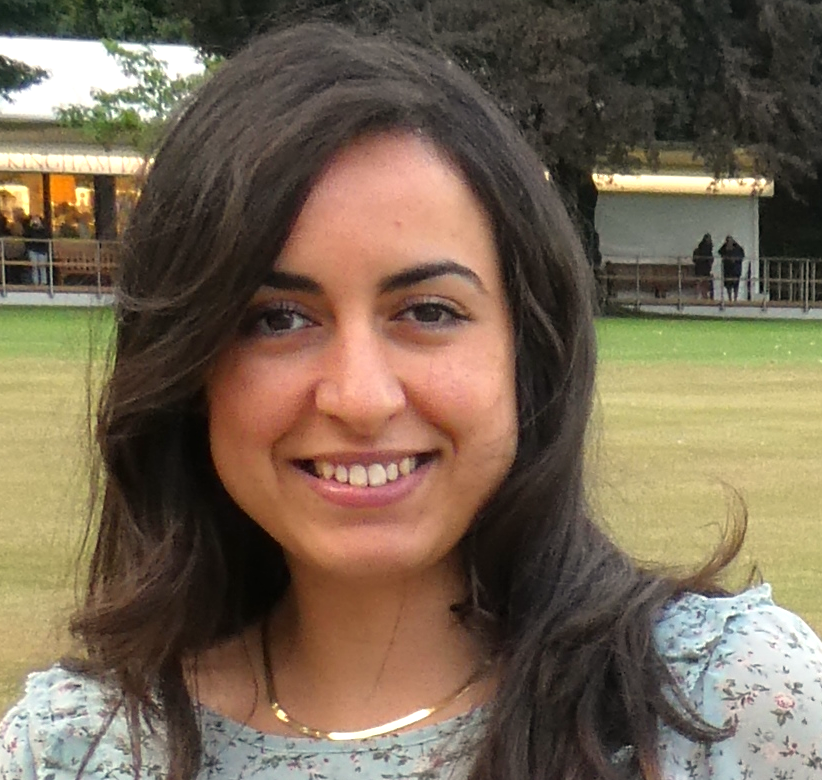}}]
{Amina Boubendir} is a researcher and project manager at Orange Labs in France. Her research is in the area of design, management and softwarization of networks and services. Amina received a Master degree in Network Design and Architecture from Télécom Paris in 2013, and a PhD in Networking and Computer Science from Télécom Paris in 2016. She is a member of the Orange Expert community on "Networks of Future".
\end{IEEEbiography}

\begin{IEEEbiography}[{\includegraphics[width=1in,height=1.25in,clip,keepaspectratio]{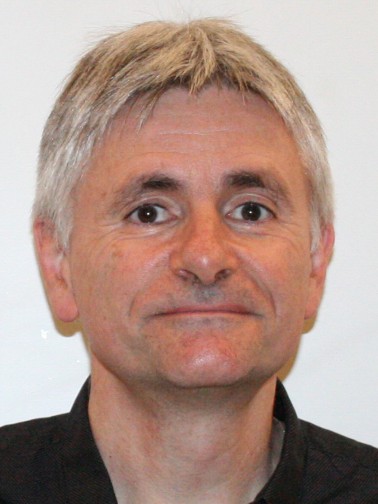}}]
{Fabrice Guillemin}
graduated from Ecole Polytechnique in 1984 and from Telecom Paris in 1989. He received the PhD degree from the University of Rennes in 1992. He defended his “habilitation” thesis in 1999 at the University Pierre et Marie Curie (LIP6), Paris. Since 1989, he has been with Orange Labs (former CNET and France Telecom R\&D). He is currently leading a project on the evolution of network control. He is a member of the Orange Expert community on "Networks of Future".
\end{IEEEbiography}


\begin{IEEEbiography}[{\includegraphics[width=1in,height=1.25in,clip,keepaspectratio]{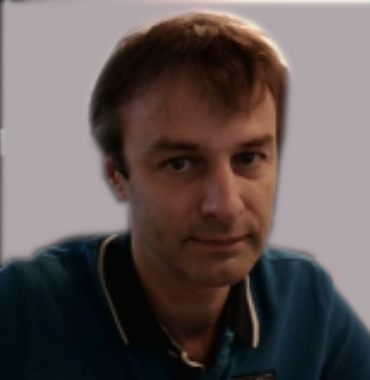}}]{Pierre Sens}
received his Ph.D. in Computer Science in 1994, and the “Habilitation à diriger des recherches” in 2000 from Paris 6 University (UPMC), France. Currently, he is a full Professor at Sorbonne Université (ex-UPMC). His research interests include distributed systems and algorithms, large scale data storage, fault tolerance, and cloud computing. He is leading Delys a joint research team between LIP6 and Inria. He was member of the Program Committee of major conferences in the areas of distributed systems and parallelism (ICDCS, IPDPS, OPODIS, ICPP, Europar, SRDS, DISC. . . ) and has served as general chair of SBAC-PAD and EDCC. Overall, he has published over 150 papers in international journals and conferences and has acted for advisor of 25 Ph.D. thesis.
\end{IEEEbiography}



\end{document}